\documentclass[reprint,superscriptaddress,amsmath,amssymb,aps, pra]{revtex4-1}

\usepackage[T1]{fontenc} 

\usepackage{graphicx}
\usepackage{dcolumn}
\usepackage{bm}

\usepackage{lipsum}
\usepackage{color}
\usepackage[normalem]{ulem}
\usepackage{simplewick}
\usepackage{qcircuit}
\usepackage{braket}
\usepackage{float}
\usepackage{multirow}
\usepackage{tikz}
\usepackage[colorlinks=true,urlcolor=blue,citecolor=blue,linkcolor=blue]{hyperref}

\begin{document}

\title{Quantum algorithms for electronic structure calculations: particle/hole Hamiltonian and optimized wavefunction expansions}

\author{Panagiotis Kl. Barkoutsos}
\affiliation{IBM Research GmbH, Zurich Research Laboratory, S\"aumerstrasse 4, 8803 R\"uschlikon, Switzerland}
\affiliation{Institute for Theoretical Physics, ETH Zurich, 8093 Zurich, Switzerland}

\author{Jerome F. Gonthier}
\affiliation{Kenneth S. Pitzer Center for Theoretical Chemistry, Department of Chemistry,University of California, Berkeley, CA 94720, USA}

\author{Igor Sokolov}
\affiliation{IBM Research GmbH, Zurich Research Laboratory, S\"aumerstrasse 4, 8803 R\"uschlikon, Switzerland}
\affiliation{Institute for Theoretical Physics, ETH Zurich, 8093 Zurich, Switzerland}

\author{Nikolaj Moll}
\affiliation{IBM Research GmbH, Zurich Research Laboratory, S\"aumerstrasse 4, 8803 R\"uschlikon, Switzerland}

\author{Gian Salis}
\affiliation{IBM Research GmbH, Zurich Research Laboratory, S\"aumerstrasse 4, 8803 R\"uschlikon, Switzerland}

\author{Andreas Fuhrer}
\affiliation{IBM Research GmbH, Zurich Research Laboratory, S\"aumerstrasse 4, 8803 R\"uschlikon, Switzerland}

\author{Marc Ganzhorn}
\affiliation{IBM Research GmbH, Zurich Research Laboratory, S\"aumerstrasse 4, 8803 R\"uschlikon, Switzerland}

\author{Daniel J. Egger}
\affiliation{IBM Research GmbH, Zurich Research Laboratory, S\"aumerstrasse 4, 8803 R\"uschlikon, Switzerland}

\author{Matthias Troyer}
\affiliation{Institute for Theoretical Physics, ETH Zurich, 8093 Zurich, Switzerland}
\affiliation{Microsoft Quantum, Microsoft, Redmond, WA 98052, USA}

\author{Antonio Mezzacapo}
\affiliation{IBM T.J. Watson Research Center, Yorktown Heights, NY 10598, USA}

\author{Stefan Filipp}
\affiliation{IBM Research GmbH, Zurich Research Laboratory, S\"aumerstrasse 4, 8803 R\"uschlikon, Switzerland}

\author{Ivano Tavernelli}
\email{ita@zurich.ibm.com}
\affiliation{IBM Research GmbH, Zurich Research Laboratory, S\"aumerstrasse 4, 8803 R\"uschlikon, Switzerland}

\date{\today}

\begin{abstract}
In this work we investigate methods to improve the efficiency and scalability of quantum algorithms for quantum chemistry applications. 
We propose a transformation of the electronic structure Hamiltonian in the second quantization framework into the \textit{particle-hole (p/h)} picture, which offers a better starting point for the expansion of the trial wavefunction.
The state of the molecular system at study 
is parametrized in a way to efficiently explore the sector of the molecular Fock space that contains the desired solution.
To this end, we explore several trial wavefunctions
to identify the most efficient parameterization of the molecular ground state.
Taking advantage of known post-Hartree Fock quantum chemistry approaches and heuristic Hilbert space search quantum algorithms, we propose a new family of quantum circuits based on exchange-type gates
that enable accurate calculations while keeping the gate count (i.e., the circuit depth) low.
The particle-hole implementation of the Unitary Coupled Cluster (UCC) method within the Variational Quantum Eigensolver approach gives rise to an efficient quantum algorithm, named \textit{q-UCC}, with important advantages compared to the straightforward `translation' of the classical Coupled Cluster counterpart. In particular, we show how a single Trotter step can accurately and efficiently reproduce the ground state energies of simple molecular systems.

\end{abstract}

\pacs{Valid PACS appear here}
\maketitle

\section{\label{sec:level1}Introduction}

Quantum computing is emerging as a new paradigm for the solution of a wide class of problems that are not accessible by conventional high performance computers based on classical algorithms \cite{lloyd_universal_1996, preskill18}. 
Quantum computers can in principle efficiently solve problems that require exponential resources on classical hardware, even when using the best known classical algorithms.
In the last few years, several interesting problems with potential quantum speedup have been brought forward in the domain of quantum physics, like 
eigenvalue-search using quantum phase estimation algorithms~\cite{Nielse_Chuang2011,abrams_quantum_1999, Brassard2000, Hoyer2000} 
and evaluation of observables in quantum chemistry \cite{abrams_simulation_1997, lanyon_towards_2010,Du2010, Wang_quantum_2015, Mueck2015, reiher_elucidating_2017, shen_quantum_2017}, e.g. by means of the hybrid variational quantum eigensolver (VQE) algorithm~\cite{peruzzo_variational_2014, Yung2014, barends_digital_2015, mcclean_theory_2016,  wang18}.

The original idea that a quantum computer can potentially solve many-body quantum mechanical problems more efficiently than classical algorithms is due to R.~Feynman who proposed to use quantum algorithms to investigate the fundamental properties of nature at the quantum scale~\cite{Feynman1965, Feynman1982},
while there are still no classical algorithms with favourable scaling that find the `exact' solution of quantum mechanical problems.
Using different systematic expansions of the many-electron wavefunction,
several quantum chemistry approaches have been proposed that can reach 
an arbitrary precision for the ground state energy of the molecular Hamiltonian \cite{Rubin2018, Kivlichan2018, Babbush2018}.
The most commonly used variational approaches are 
full Configuration Interaction (full CI)~\cite{Sherrill1999} 
and Coupled Cluster (CC)~\cite{Bartlett2007}.
However, for all these approaches the scaling as a function of the number of degrees of freedom $N$ (e.g., number of electrons or number of basis functions) is very unfavorable: $\mathcal{O}(N!)$ in full CI~\footnote{The correct scaling will be $\begin{pmatrix} N_{b} \\ N_{\rm el}\end{pmatrix}$, where $N_{b}$ is the number of basis functions and $N_{\rm el}$ is the number of electrons.}
and $\mathcal{O}(N^{10})$ for the CC approach when the expansion is truncated at the fourth order in the electronic excitation operator, named CCSDTQ (S stands for single, D for double, T for triple and Q for quadruple excitations).
At present, the CCSD(T) expansion (that includes an approximated treatment of the triples excitations~\cite{Raghavachari1989, Bartlett2007}) with a scaling $\mathcal{O}(N^7)$ is often considered the `gold standard' for quantum chemistry calculations. 
Energies computed at CCSD(T) level of theory have an error that lies within the so-called \textit{chemical accuracy} (errors less than 1~-~5~kcal/mol = 0.043~-~0.22~eV) for many systems (i.e., when no strong static correlation or multi-reference character of the ground state is present~\cite{Becke2013, Ziesche1997}).
The exponential scaling of Hilbert space as function of the number of qubits in quantum computers opens up new possibilities for the calculation of accurate electronic structure properties using quantum devices.

Designing quantum algorithms for quantum chemistry calculations requires reformulating the fermionic problem into qubit operators.
This includes 
(\textit{i}) the mapping of the original  electronic structure Hamiltonian into the corresponding qubit Hamiltonian; 
(\textit{ii}) the preparation of suitable trial wavefunctions,  
and 
(\textit{iii}) the development of an optimization scheme that converges to a ground state solution compatible with the nature of the quantum circuit.
As for the mapping (\textit{i}), we will work in the second quantization formalism (SQ) of quantum mechanics. 
The main reason for this choice is that the degrees of freedom are encoded in the expansion coefficients of the electronic wavefunction. 
This avoids the costly discretization of the physical space needed  
in the first quantization (FQ) picture.
The SQ approach has the clear advantage of being readily applicable to small molecular systems using state-of-the-art quantum architectures, while methods in FQ will require a larger number of qubits even for the simulation of small systems such as $\rm H_2$.

The SQ Hamiltonian is formulated in the Hartree-Fock (HF) basis and mapped to the qubit space using either the Jordan-Wigner~\cite{jordan_uber_1928}, the Bravyi-Kitaev~\cite{bravyi_fermionic_2002} or the parity mapping transformations~\cite{bravyi_tapering_2017}.
This formalism was already successfully applied to the study of a number of small size molecular systems, from molecular Hydrogen~\cite{omalley_scalable_2016,kandala_hardware-efficient_2017, Colless2018, Hempel2018}, $\rm{H_2}$, to Beryllium dihydrate,  $\rm{BeH_2}$~\cite{kandala_hardware-efficient_2017}.
While the scaling of this approach is not yet fully understood, the complexity of the problem can be reduced by the 
encoding of specific symmetries directly at the Hamiltonian level.
For example, one can restrict the action of the SQ Hamiltonian to the sector of the Fock space that corresponds to the desired number of electrons~\cite{moll_optimizing_2016} or implement symmetry constraints~\cite{bravyi_tapering_2017}.

The trial wavefunction (\textit{ii}), can be prepared with either of two main strategies.
First, one can \textit{translate} classical approaches (full CI, CC, and alike) in the qubit language by designing circuits parametrized in the angles of single and two-qubit gates.
This method (that we name classically inspired approach, CLA) was pioneered by several research groups worldwide~\cite{whitfield_computational_2012, Seeley2012, romero_strategies_2017} using the CC Ansatz truncated at different levels of excitations.
This approach suffers from different drawbacks, e.g. the number of parameters (gate angles) increases significantly with the number of electrons, impacting seriously the efficiency of the parameter optimization and limiting therefore the scaling to larger systems.
The second approach, named heuristic sampling~\cite{kandala_hardware-efficient_2017}, prepares the trial state using single qubit rotations and hardware efficient entangler blocks that span the whole qubit register.
This heuristic approach (HEA) does not have any equivalent `classical' counterpart since it was designed to exploit the unique capabilities of the quantum hardware.
In both cases (CLA and HEA trial wavefunctions), the optimization of the parameters, point (\textit{iii}), is done using a classical optimization algorithm (e.g. the Simultaneous Perturbation Stochastic Approximation (SPSA) algorithm~\cite{spall_adaptive_2000}). 
The overall approach falls therefore into the class of the Variational Quantum Eigensolver (VQE) algorithm, where the exponentially hard part of the problem (the sampling of the wavefunction space) and the calculation of the Hamiltonian expectation values are performed in the quantum hardware, while the parameter optimization is done in a classical computer.

The paper is organized as follows. 
In Section~\ref{section_theory}, we discuss the mapping of the SQ Hamiltonian into the particle-hole picture. 
To keep a one-to-one correspondence with the classical UCCSD algorithm we do not perform any additional reduction of the Hamiltonian as done in previous studies~\cite{moll_optimizing_2016}. 
One of the aims of this work is in fact to investigate the relations between the classical CCSD and the quantum UCCSD algorithms to identify possible strategies for a more efficient implementation of the CC  expansion in quantum circuits.
The possibility to apply specific parametrized particle-conserving exchange-type gates in the heuristic approach is also discussed.
Section~\ref{subsec:subsection_vqe_algorithm} discusses the implementation of the VQE algorithm in the particle-hole formalism.
In Section~\ref{section_results}, we apply these techniques to 
the hydrogen (${\rm H_2}$) and water (${\rm H_2O}$) molecules 
and discuss the impact of the different approximations. 
Conclusions are summarized in Section~\ref{section_conclusions}.

\section{THEORY}
\label{section_theory}
The \textit{particle-hole (p/h)} representation~\cite{Fetter2003} provides a better reference trial wavefunction that improves the performance of the VQE optimization algorithm.
The optimization in the particle-hole framework is performed using two different trial wavefunction Ans\"atze: 
the CC-based expansion~\cite{romero_strategies_2017} and 
the heuristic approach~\cite{kandala_hardware-efficient_2017}. 
To improve the efficiency and scalability of these methods we investigate different approximations and their associated errors.  

\subsection{Hamiltonian in the particle-hole picture}
\label{subsec_ph_Hamiltonian}
We start with the electronic structure SQ Hamiltonian in the Hartree-Fock orbitals basis $\{\phi_i(r)\}_{i=1}^{N_{\text{max}}}$~\cite{szabo12}, 
\begin{equation}
\hat H^{el} = \sum_{ij} h_{ij} \,  \hat{a}^{\dagger}_i \hat{a}_j +
\sum_{ijkl} g_{ijkl} \,   \hat{a}^{\dagger}_i \hat{a}^{\dagger}_j \hat{a}_l \hat{a}_k
\label{Eq:H_sec_quant}
\end{equation}
where $h_{ij} = \langle i |\hat h| j \rangle$ are the one-electron integrals defined as
\begin{equation}
 \langle i |\hat h| j \rangle = \int dr_1 \, \phi_i^*(r_1) \left(-\frac{1}{2} \nabla^2_{r_1} - \sum^M_{I=1} \frac{Z_I}{R_{1I}}\right) \phi_j(r_1) \label{eq:hij} 
\end{equation}
and $g_{ijkl} = \langle i j |\hat g| k l \rangle$ the two-electron terms given by
\begin{equation}
\langle i j |\hat g| k l \rangle = \int d r_1 dr_2 \, \phi_i^*(r_1) \phi_j^*(r_2) \frac{1}{r_{12}} \phi_k(r_1) \phi_l(r_2) \, . \label{eq:gijkl}
\end{equation}
Here $R_I, r_i\in R^3$ are the coordinates of atom $I$ and electron $i$, respectively.
In Eq.~\eqref{eq:hij} $M$ is the total number of atoms in the system, $Z_I$ are the atomic numbers, $\nabla^2_{r_1}=\partial^2_{x_1} + \partial^2_{y_1} + \partial^2_{z_1}$, $R_{1I}=|r_1-R_I|$, and $r_{12}=|r_1-r_2|$. 
Throughout the paper we use the `physicists' notation for the definition of the two-electron integrals~\cite{szabo12}. 
The Hamiltonian in Eq.~\eqref{Eq:H_sec_quant} acts in the Fock space $\mathcal{F}=\bigoplus_{N=0}^{N_{\rm  max}} \mathcal{A} \mathcal{H}^{\otimes N}$ with particle number $N\in \{0, \dots N_{\rm max}\}$, where $\mathcal{H}$ is the one-particle Hilbert space and $\mathcal{A}$ the anti-symmetrizing operator.

To move to the p/h representation, we start with the definition of a new vacuum state in the $N$-particle sector of the Fock space
\begin{equation}
|\Phi_0 \rangle = \prod_{i=1}^N \hat{a}^{\dagger}_i|\text{vac} \rangle \, ,
\label{Eq:HF_GS}
\end{equation}
which coincides with the Slater determinant solution of the HF problem with $N$ electrons. The set of HF orbitals contributing to $|\Phi_0 \rangle$ are called occupied $\{\phi_i(r)\}_{i=1}^N$, while all others high energy orbitals are called unoccupied or virtual, $\{\phi_i(r)\}_{i=N+1}^{N_{\text{max}}}$. 
In this work, we will use the following notation for the orbital indices: $i,j,k,l$: for occupied orbitals; $m, n, p, q$: for virtual (unoccupied) orbitals; $r,s,t,u$ for either types.
A generic state can then be generated from the new ground state $|\Phi_0 \rangle $ using excitation operators that create \textit{holes} within the set of occupied orbitals and \textit{particles} within the unoccupied or virtual set. 
For instance, the excitation operator $\hat{a}^{\dagger}_m \hat{a_i}$ excites one electron from the occupied HF orbital $\phi_i(r)$ into the unoccupied orbital $\phi_m(r)$.
The \textit{holes} and \textit{particles} generated by the excitation operators with respect to the ground state $|\Phi_0 \rangle$ are called \textit{quasi-particles}.
The corresponding creation and annihilation operators are defined by
\begin{align}
&\hat b_i^{\dagger} =  \hat{a}_i & \quad  \text{(hole creation)} \label{eq:def_b_start}\\
& \hat b_m^{\dagger}  = \hat{a}^\dagger_m & \quad \text{(particle creation)} \\
& \hat b_i = \hat{a}^\dagger_i & \quad  \text{(hole annihilation)} \\
&\hat b_m = \hat{a}_m & \quad  \text{(particle annihilation)} \label{eq:def_b_end}
\end{align}
and still fulfill the fermionic anti-commutation relation statistics.
In the `quasi-particle' framework we can define a normal ordering operator $\hat{N}_b[\dots]$.
With $\hat{N}_b$ we define an equivalent electronic structure Hamiltonian in the \textit{particle-hole} (p/h) picture that has $|\Phi_0 \rangle$ as reference (vacuum) state.
This Hamiltonian is
\begin{align}
\hat{H}^{p/h}=&E_{HF} + 
\sum_{rs} \langle r|\hat F|s\rangle \hat{N}_b[\hat a_r^{\dagger} \hat a_s]  \notag \\
&+\frac{1}{2} \sum_{srtu} \langle r s |\hat g| t u\rangle \hat{N}_b[\hat a_r^{\dagger} \hat a^{\dagger}_s \hat a_u \hat a_t]
\label{eq:HF_H_ph}
\end{align}
where
\begin{equation}
E_{HF} = \sum_i \langle i | \hat h | i \rangle + \frac{1}{2} \sum_{ij} (\langle ij | \hat g | ij \rangle-\langle ij | \hat g | ji \rangle ) 
\end{equation}
is the reference energy and $\langle r|\hat F|s\rangle$ is the Fock matrix
\begin{equation}
\langle r|\hat F|s\rangle = \langle r | \hat h | s \rangle + \sum_{i} (\langle ri | \hat g | si \rangle-\langle ri | \hat g | is \rangle ) \, .
\end{equation}
In Eq.~\eqref{eq:HF_H_ph}, the normal ordering operator $\hat{N}_b$ acts on the p/h operators $\{\hat{b}_r, \hat{b}^{\dagger}_s\}$, which appear after applying the transformations in Eq.~\eqref{eq:def_b_start}-\eqref{eq:def_b_end}.

The advantage of this transformation is evident if we think about perturbation theory applied to the ground state in Eq.~\eqref{Eq:HF_GS}.
Only after redefining the normal ordering as in Eqs.~\eqref{eq:def_b_start}-\eqref{eq:def_b_end} it is possible to obtain an efficient perturbative expansion using Wick's theorem, which independent of the number of electrons in the system.
Note that the transformation to the p/h picture can be obtained by applying a rotation to the HF ground state or by performing the transformation described above leading to the p/h Hamiltonian in Eq.~\eqref{eq:HF_H_ph}.
For practical convenience, we chose the second approach as in the VQE algorithm the Hamiltonian is a measured quantity while the wavefunction is encoded in the qubit register and therefore it should be kept in its original form~\footnote{To implement the UCCSD wavefunction Ansatz, we expand the Hamiltonian in the basis function of the occupied and virtual HF orbitals, with a number of occupied orbitals equal to the number of electrons in the system. 
This picture has the advantage of allowing a simple interpretation of the expansion of the reference wavefunction in terms of excited configurations (Slater determinants)
Further manipulations of the molecular Hamiltonians in the unmodified second quantized form (Eq.~\eqref{Eq:H_sec_quant}) or in the p/h formulation can be used to further reduce the number of required qubits. One possibility, is to apply the projection scheme introduced in~\cite{moll_optimizing_2016, Barkoutsos_fermionic_2017}, which allow to restrict the search space from the entire Fock space to the sector of the Hilbert space with the selected number of electrons.
However, this procedure will make the physical interpretation of the UCC expansion less evident and the mapping to the quantum circuits more cumbersome.
For these reasons, in this work we will restrict to the simplest map that encodes each basis function in a different qubit.}.

\subsection{Trial wavefunctions}
\label{subsection_WF}
The trial wavefunctions are constructed applying a set of perturbations (`excitations') to the HF ground state wavefunction, $|\Phi_0 \rangle$. 
The perturbations are controlled by a set of parameters (gate angles) that are then optimized until convergence is reached.

We can identify two main classes of trial wavefunctions:
The first one, based on the CC Ansatz, provides a controllable and intuitively simple expansion of the initial HF wavefunction combined with an efficient parameterization of the final state, minimizing therefore the number of independent parameters.
The unitary version of the CC approach (UCC~\cite{Bartlett1989}), is more suited for applications in quantum computing due to the properties of the applied gate operations.
While often implemented as a variational approach, UCC still differs from the truly variational version of CC (vCC)~\cite{Kutzelnigg1991}.
However, the difference between the UCC and variational-CC energies is in general very small~\cite{Harsha2018}. 

The second class of trial wavefunctions is based on quantum algorithms that have no strict classical equivalent. 
In fact, these approaches are not based on a controlled perturbative expansion around a zero-order solution (e.g., the HF state) but instead they aim at sampling in the most efficient way possible the relevant portion of the Hilbert space that contains the solution.

\subsubsection{The UCC Ansatz} 
\label{sss:UCC}
In UCC the trial wavefunction is parametrized using the following Ansatz
\begin{equation}
|\Psi(\vec{\theta})\rangle =
e^{\hat T(\vec{\theta})- \hat T^{\dagger}(\vec{\theta})} |\Phi_0\rangle
\label{eq:UCCmain}
\end{equation}
where $\hat T(\vec{\theta})=\hat T_{1}(\vec{\theta})+\hat T_{2}(\vec{\theta}) + \dots + \hat T_{n}(\vec{\theta})$ is the excitation operator to order $n$ with
\begin{align}
&\hat T_{1}(\vec{\theta})=\sum_{i;m}\theta^{m}_{i} \, \hat a^{\dagger}_m \hat a_i \label{eq:t1} \\
& \hat T_{2}(\vec{\theta})=\frac{1}{2} \sum_{i,j; m,n}\theta^{m,n}_{i,j} \, \hat a^{\dagger}_n \hat a^{\dagger}_m \hat a_j \hat a_i
\label{eq:t2}
\end{align}
(with equivalent expressions for higher orders) and
$\vec{\theta}=\{ \{\theta^{m}_{i}\},\{\theta^{m,n}_{i,j}\},\dots \}$ is a collective vector for all expansion coefficients.
In the following, we will restrict our implementation to the UCCSD case i.e, $\hat T=\hat T_{1}(\vec{\theta})+\hat T_{2}(\vec{\theta})$,  the extension to higher excitations does not imply any further development but only requires the implementation of longer circuits that are at present unpractical from experimental, as well as simulation perspectives.

The correlation energy of the system (i.e., the correction to the HF energy) is given by $\langle \Psi(\vec{\theta}) | \hat H^{p/h}|\Psi(\vec{\theta})\rangle -E_{HF}$ using the p/h Hamiltonian of Eq.~\eqref{eq:HF_H_ph}. 
The VQE algorithm will find the optimal $\vec{\theta}$ parameters from which the correlated ground state energy is obtained
\begin{equation}
E_{GS}= E_{HF} + E^{\text{corr}}_{UCCSD}(\vec{\theta}_{\text{min}})
\end{equation}
where $E_{HF}=\langle \Phi_0 | \hat H^{el} |\Phi_0 \rangle$.

\begin{figure}[h]
    \centering
    
    \begin{tikzpicture}
    \node at (0,0.25) {\includegraphics[width = 0.8\columnwidth]{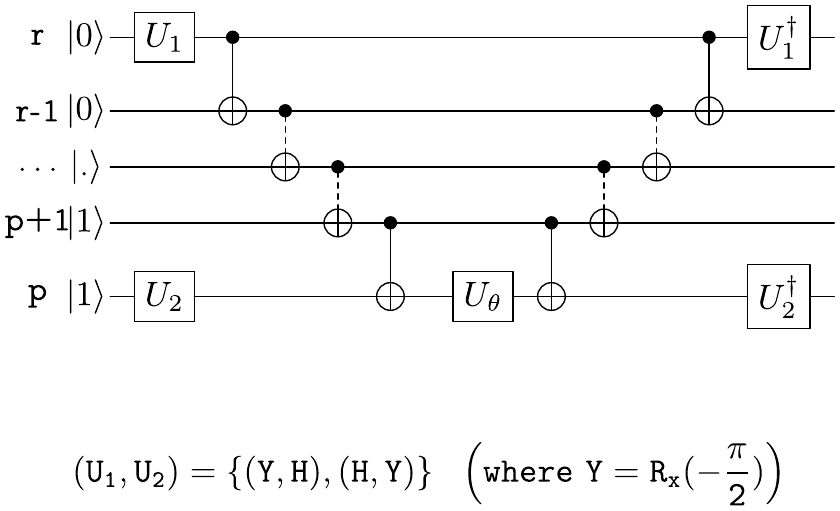}};
    \node at (-3.8,2.5) {(a)};
    \node at (0,-5.25) {\includegraphics[width = 0.85\columnwidth]{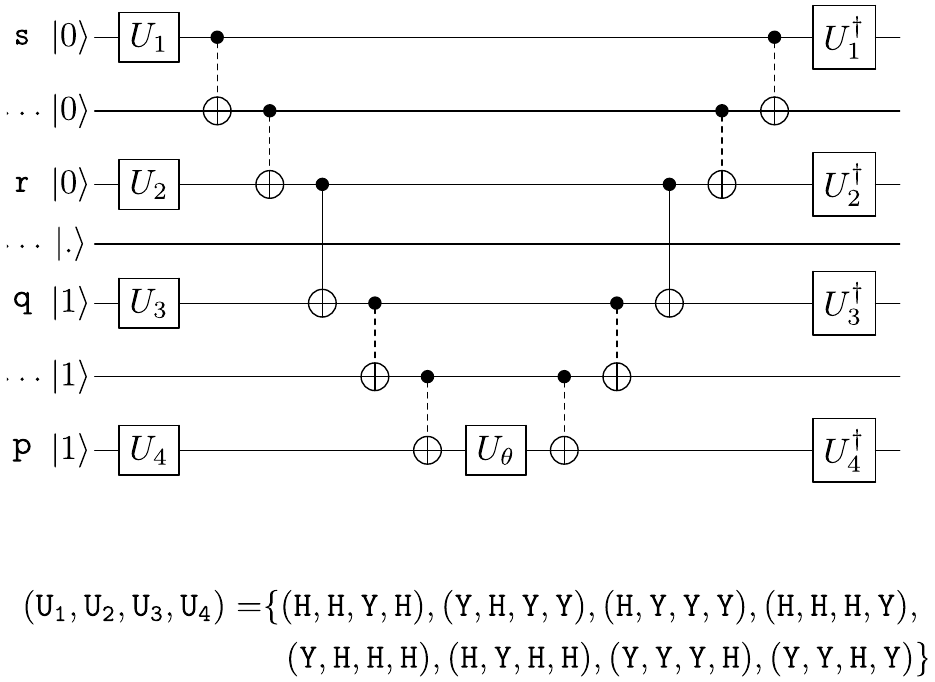}};
    \node at (-3.8,-2.5) {(b)};
    \end{tikzpicture}
    \caption{
    Circuits for the exponentiation of the single (a)
    and double (b) excitation operators 
    $(\hat a_{p}^{\dag} \hat a_r - h.c.)$ 
    and 
    $(\hat a_p^{\dag} \hat a_q^{\dag} \hat a_r \hat a_s - h.c.)$, 
    which contribute to $\hat T_{1}$
    and $\hat T_{2}$, respectively.
    The $p$, $q$ indices refer to virtual and $r$,$s$ to occupied orbitals. The generic state $| .  \rangle$ corresponds to $| 1 \rangle$ in case it is part of the occupied manifold and $| 0 \rangle$ otherwise. 
    The repeated units across several qubits are shown in dashed lines. The definition of the gates that span more than two qubits (dashed lines) is given in Appendix~\ref{App:boxes}.}
    \label{fig:UCCSD_circ_T1}
\end{figure}

The circuits for the implementation of the UCCSD trial wavefunction (see Fig.~\ref{fig:UCCSD_circ_T1}) are constructed following the prescriptions in~\cite{Nielse_Chuang2011, whitfield_computational_2012, romero_strategies_2017} and implemented in the IBM software package QISKit~\cite{qiskit_webpage}. 

Particularly challenging for the implementation of UCCSD expansion is the mapping to circuits of the exponentiation in Eq.~\eqref{eq:UCCmain}, which results in a circuit depth that scales as 
$\mathcal{O}\left( \binom{N_{\rm occ}}{2} \times \binom{N_{\rm virt}}{2} \times N_{\rm qubits}
\right)$, 
where $N_{\rm occ}$ ($N_{\rm vir}$) is the number of occupied (virtual) orbitals that take part to the excitations.
In this work, we therefore investigate the impact of some approximations used for the implementation of the UCC expansion in quantum circuits.
In particular, we will focus on the effect of applying an increasing number of Trotter steps, $n$, in the approximation of the expansion
\begin{equation}
e^{(\hat A+ \hat B)} = \lim_{n \rightarrow \infty}\left(
e^{\frac{\hat A}{n}} e^{\frac{\hat B}{n}}
\right)^n \, ,
\label{eq_trotter}
\end{equation}
in the case of non-commuting operators $\hat A$ and $\hat B$. 
This situation occurs for instance when dealing with terms of the form $\hat A=\theta_{i}^{m} (\hat a_i^{\dagger} \hat a_m - \hat a_m^{\dagger} \hat a_i )$ and $\hat B=\theta_{ij}^{mn} (\hat a^{\dagger}_i \hat a^{\dagger}_m \hat a_k \hat a_n-  \hat a^{\dagger}_n \hat a^{\dagger}_k \hat a_m \hat a_i)$. 

Finally, it is important to stress that the classical (non-variational) CCSD approach~\cite{Bartlett2007} leads in fact to a set of closed equations for the parameters in Eqs.~\eqref{eq:t1} and~\eqref{eq:t2} by fully exploiting the commutation relation of the $\hat T_1$ and $\hat T_2$ operators and the properties of the normal ordering operator. 
However, these properties are not applicable in the `variational' UCCSD formulation due to the replacement of the $\hat T_i$ by the  $(\hat T_i-\hat T_i^\dagger)$ operators for $i=1,2$.
More details on the approximations used in the UCCSD approach are summarized in Appendix~\ref{section:appendix_UCCSD_approx}.

\subsubsection{The Heuristic Ansatz}
\label{sss:Heuristic}

In addition to the UCC Ansatz, 
in this work we also investigate the adjustment of the HEA approach to the p/h formalism.
In particular, 
we design particle conserving entangler blocks to constrain the wavefunction search to the sector of Hilbert space with a constant number of particles and 
we investigate the advantage of using hardware specific exchange-type gates~\cite{McKay2016, Roth2017, Egger2018}.
The preparation of the heuristic trial states comprises two types of quantum operations, single-qubit Euler rotations $\hat U(\vec{\theta})$ with angles $\vec{\theta}$ and an entangling \emph{drift} operation $\hat U_{\rm{ent}}(\vec{\theta})$ acting on pairs of qubits. 
The $N$-qubit trial states are obtained by applying a sequence of $D$ entanglers $\hat U_{\rm ent}$ 
alternating with the Euler rotations on the $N$-qubits to the HF ground state $|\Phi_0 \rangle$,
\begin{equation}
| \Psi(\vec{\theta}) \rangle = \overbrace{ 
\hat U^{D}(\vec{\theta}) \hat U_{\rm ent}\ldots \hat U^{1}(\vec{\theta}) \hat U_{\rm ent}}^{\rm{D-times}} \hat U^{0}(\vec{\theta}) |\Phi_0 \rangle
\label{eq:entangler}
\end{equation}
The choice of the initial HF state $|\Phi_0 \rangle$ improves the efficiency of the searching algorithm, avoiding Barren plateaus in Hilbert space~\cite{McClean_2018_Baren}.

This gate sequence has $p = N (3 D + 2)$ independent angles.
In full analogy with our UCC approach, we make use of the particle-hole  Hamiltonian $\hat H^{p/h}$ expressed in terms of the HF orbitals instead of the original Hamiltonian in second quantization (Eq.~\eqref{Eq:H_sec_quant}) as done in~\cite{kandala_hardware-efficient_2017}.
Within this framework the most suited entangler blocks are made up of particle-conserving gates, i.e. gates that conserve the number of excited qubits.

More specifically, the single-qubit operations are decomposed into rotations about the $x-$ and the $z-$axes, $ \hat{U}^{q,k}(\vec{\theta}) = \hat Z^q_{\theta^{q,k}_1} \hat X^q_{\theta^{q,k}_2} \hat  Z^q_{\theta^{q,k}_3}$, 
where 
\begin{equation}
\hat X^q(\theta^{q,k}_j) = \exp\left[-i\theta^{q,k}_j\hat{\sigma}^x_q/2\right]
\end{equation}
denotes the unitary operation acting on qubit $q$ at the $i$-th position of the gate sequence (similarly for $\hat{Z}^q(\theta^{q,k}_j)$) \cite{murnaghan62, Nielse_Chuang2011}. 

\begin{figure}[h]
    \centering
    \begin{tikzpicture}
    \node at (0,0) {\includegraphics[width = 0.91\columnwidth]{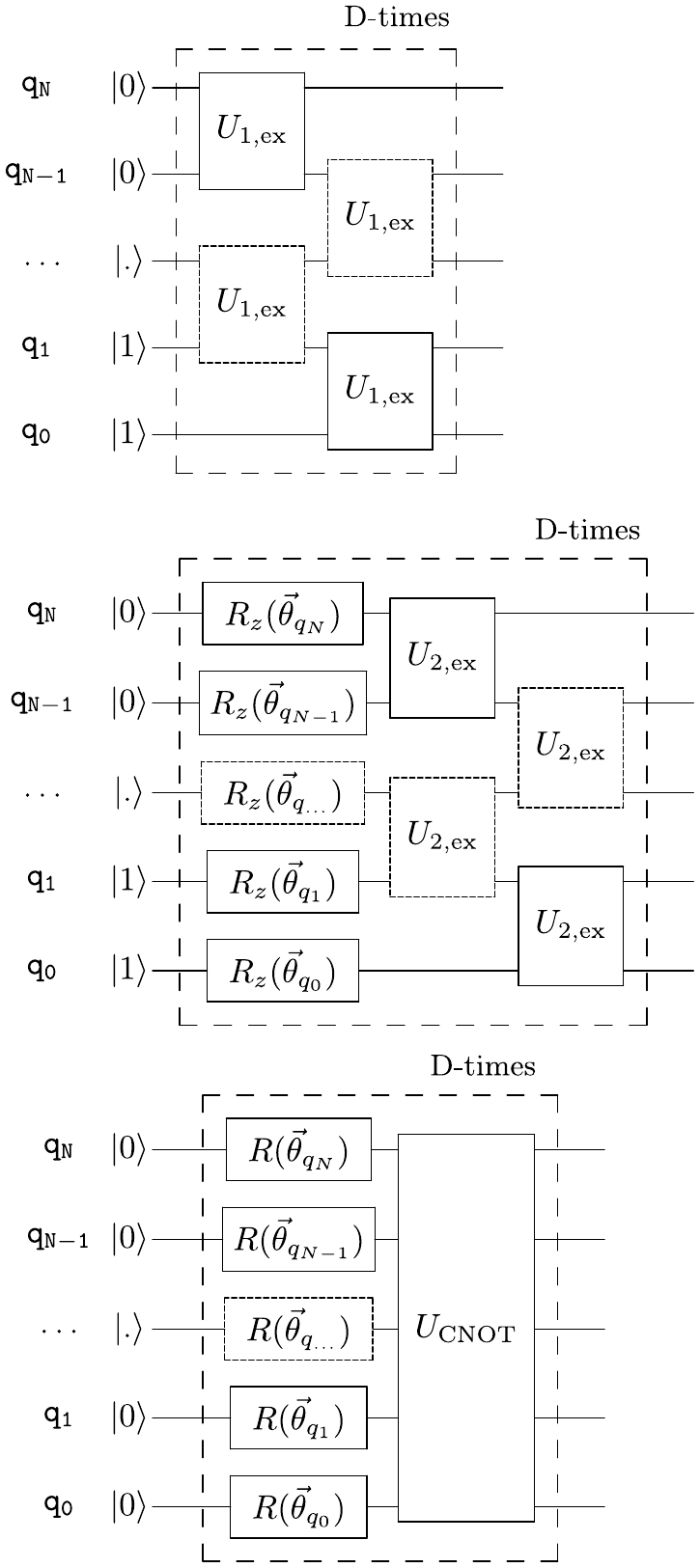}};
    \node at (-3.65,8.5) {(a)};
    \node at (-3.65,2.6) {(b)};
    \node at (-3.65,-3.2) {(c)};
    \end{tikzpicture}
    \caption {Definition of the three entangler blocks: (a) ${U}^{(1)}_{\rm ent}$, (b) ${U}^{(2)}_{\rm ent}$ and (c) ${U}^{(3)}_{\rm ent}$, composed by the ${U}_{1,\rm{ex}}$
    , see Eq.~\eqref{Eq:USWAP}, 
    ${U}_{2,\rm{ex}}$, see Eq.~\eqref{Eq:UFLIP} and CNOT
    gates, respectively. 
    The repeated units across several qubits are shown in dotted boxes
    (see Appendix~\ref{App:boxes}).
    }
    \label{fig:circuits}
\end{figure}
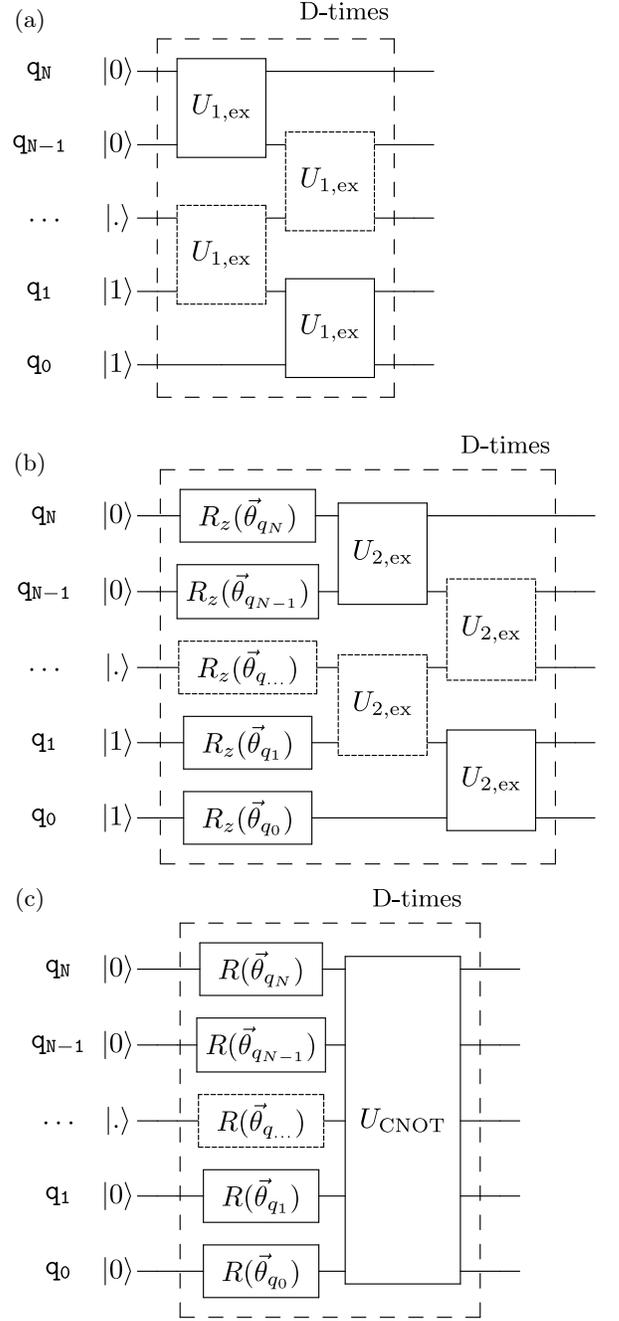

In this work, we investigate the performance of three different 
entangler blocks, $U^{(1-3)}_{\rm ent}$ (Fig.~\ref{fig:circuits}), composed by:
(1) the particle conserving two-parameter exchange-type gate, defined by 

\begin{equation}
   U_{\text{1,ex}} (\theta_1,\theta_2) = \begin{pmatrix} 
      1 & 0 & 0 & 0 \\
      0 &\cos \theta_1 & e^{i \theta_2} \sin \theta_1 & 0\\
      0 & e^{-i \theta_2} \sin \theta_1 & -\cos \theta_1 & 0\\
      0 & 0 & 0 & 1
    \end{pmatrix}
    \label{Eq:USWAP}
\end{equation}
parametrized by the angles $\theta_1$ and $\theta_2$ \cite{Egger2018}, 
(2) the particle conserving single-parameter exchange-type gate

\begin{equation}
   U_{\text{2,ex}} (\theta) = \begin{pmatrix} 
      1 & 0 & 0 & 0 \\
      0 &\cos 2 \theta & -i \sin 2 \theta & 0\\
      0 &  -i \sin 2 \theta & \cos 2 \theta & 0\\
      0 & 0 & 0 & 1
    \end{pmatrix}
    \label{Eq:UFLIP}
\end{equation}
parametrized by angle $\theta$ \cite{McKay2016},
and (3) the entangler block composed by all-to-all CNOT gates, $U_{\rm CNOT}$.
Note that $U_{\text{2,ex}}$ is a special case of $U_{\text{1,ex}}$, but the entangler block associated to it (Fig.~\ref{fig:circuits}, panel b) also comprises single qubit rotations
(the decomposition of $U_{\rm{1,ex}}$ and $U_{\rm{2,ex}}$ in elementary gates is given in Appendix~\ref{App:Decomposition}).
The first two gates are, for example, capable of implementing directly in hardware a particle-conserving SWAP gate in a single step.
The structures of the three entangler blocks used in this work are shown in Fig.~\ref{fig:circuits}.
Note that in the first case there is no need to introduce one-qubit rotation gates.

The last entangler does not conserve the particle number and therefore
the optimization can explore alternative paths through regions of the Fock space that have a different number of electrons than in the initial state.
To constrain the final number of electrons to a fixed number $N$, we can add an extra potential term to the p/h Hamiltonian
\begin{equation}
    \hat H^{p/h}_{N} = \hat H^{p/h} + \mu \, (\langle \hat{N} \rangle -N)^2
    \label{eq:chem_pot}
\end{equation}
where $\hat{N}$ is the number operator and $\mu$ is a tunable parameter.
This term can be switched on gradually during the optimization to increase flexibility during the first steps of the optimization.

\subsection{Reduction of the Hilbert space}
\label{subsection_reduction}

In addition to the development of efficient methods for the generation of trial states, other strategies can be implemented to make computations more efficient.

\paragraph{Effective Core Potentials.}
The number of degrees of freedom can be reduced by replacing the inert electrons in the innermost nuclear shells of Eq.~\eqref{eq:hij} with an effective core potential given by
\begin{equation}
    h^{\rm {ECP}}_{ij} = \int dr_1 \, \phi_i^*(r_1) \left(-\frac{1}{2} \nabla^2_{r_1} - \sum^M_{I=1} V_{\rm{ECP}}(r_{1I})\right) \phi_j(r_1) \label{eq:hij_ECP}
\end{equation}
where $V_{ECP}(r_{1I})$ is defined in ref.~\cite{Alkauskas2004}.  

In practice, this allows us to restrict the number of
basis functions and consequently the number of HF orbitals (and therefore qubits) used to expand the Hamiltonians $\hat H^{el}$ (Eq.~\eqref{Eq:H_sec_quant}) and $\hat H^{p/h}$ (Eq.~\eqref{eq:HF_H_ph}).

\paragraph{Selection of the Active Space.}
In the UCC approach one can further reduce the Hilbert space in which to search  
to a subspace generated by the `reduced' operators
\begin{align}
&\hat T'_{1}(\vec{\theta})=\sum_{i';m'}\theta^{m'}_{i'} \, \hat a^{\dagger}_{m'} \hat a_{i'} \label{eq:T1_AS}\\
&\hat T'_{2}(\vec{\theta})=\sum_{i',j'; m',n'}\theta^{m',n'}_{i',j'} \, \hat a^{\dagger}_{n'} \hat a^{\dagger}_{m'} \hat a_{j'} \hat a_{i'} \label{eq:T2_AS}
\end{align}
where the indices $i',j'$ run over a subset of occupied orbitals: 
$i',j' \in \{i_{F}-N_{\rm occ}, \dots, i_F\}$, and $m',n'$ over a subset of virtual orbitals:
$ m',n' \in \{i_{F}+1, \dots,  i_{F}+1+N_{\rm vir} \} $;
$i_F$ is index of the highest occupied HF orbital, $N_{\rm occ}$ is the number of occupied and $N_{\rm vir}$ is the number of virtual orbitals included in the expansions in Eqs.~\eqref{eq:T1_AS} and~\eqref{eq:T2_AS}.
This is similar to the Complete Active Space self-consistent field (CASSCF) method used to reduce the costs of CI calculations~\cite{Jensen2007}.
The selection of the active space is often dictated by the nature of the orbitals involved in the expansion and the overlaps among them.
Using an active space in quantum algorithms 
shortens 
the overall circuit depth therefore making better use of the limited qubit coherence time \cite{unruh_maintaining_1995}.

\section{Implementation of the VQE algorithm in the p/h picture}
\label{subsec:subsection_vqe_algorithm}
Using the VQE algorithm with the p/h formalism requires:
\begin{itemize}
\item[(\textit{i})] 
calculating the HF orbitals and storing the 
needed matrix elements: $\langle i | \hat h | i \rangle$ and  $\langle ij | \hat g | ji \rangle$ using a classical computer;
\item[(\textit{ii})] 
performing a fermion-to-qubit transformation using 
the Jordan-Wigner~\cite{jordan_uber_1928, nielsen05} procedure;
\item[(\textit{iii})] 
generating the the trial wavefunctions starting from the HF ground state $|\Phi_0\rangle = | 1 1 \dots 1  00 \dots 0 \rangle$ (with $N$ `1' entries) 
by applying 
the circuits in Fig.~\ref{fig:UCCSD_circ_T1} (UCCSD approach)
and 
and Fig.~\ref{fig:circuits} (heuristic approach) to $|\Phi_0\rangle $.
In the first iteration the gate angles are chosen from a uniform distribution between $0$ and $2\pi$;
\item[(\textit{iv})] the expectation value for the p/h Hamiltonian $\hat H^{p/h}$ using the quantum computer;
\item[(\textit{v})] the energy (parametrized in the gate angles) to a classical algorithm that performs an optimization step in the parameter space and returns the updated values to point (\textit{iii});
in this work, the optimization is performed using the BFGS algorithm~\cite{fletcher70, romero_strategies_2017}.
\end{itemize}
The steps (\textit{iii}) to (\textit{v}) are iterated until convergence is reached.

\section{Results and discussion}
\label{section_results}
In this section, we report and discuss the results obtained from the application of the theoretical developments presented in Section~\ref{section_theory} on the simulation of two relatively simple molecules, hydrogen (H$_2$) and water (H$_2$O), which incorporate most of the complexity required to illustrate the efficiency of the different advancements.

All calculations are performed using the 6-31G basis set leading to a Hilbert space of dimension $2^8$ (where $8$ corresponds to the total number $N_{\text{max}}$ of HF orbitals, occupied and virtual) for the case of molecular hydrogen and of dimension $2^{12}$ for the water molecule ($N_{\text{max}}=12$).
Further, we replaced the $1s$ core electrons of oxygen with the corresponding Effective Core Potentials (ECPs), meaning that only 8 electrons are considered in the valence shell of H$_2$O.
However, as discussed in Section~\ref{subsec:Active_space_UCCSD},
we also used active spaces to further restrict the search of the ground state wavefunction to a smaller sector of the Hilbert space.

\subsection{The particle/hole Hamiltonian}

The choice of the p/h Hamiltonian introduced in Section~\ref{subsec_ph_Hamiltonian}, Eq.~\eqref{eq:HF_H_ph}, brings several important advantages compared to the plain molecular Hamiltonian in second quantization (Eq.~\eqref{Eq:H_sec_quant}).
By shifting the reference state to the HF ground state, we achieve faster convergence since the optimization only concerns the correlation energy, which in general amounts to only a few percent of the total energy.
In Table~\ref{table:Hamiltonian-comparison}, we compare the performance of the VQE algorithm for the optimization of the electronic structure of $\rm{H_2}$  based on the p/h and plain molecular Hamiltonians (Eq.~\eqref{Eq:H_sec_quant} and Eq.~\eqref{eq:HF_H_ph}, respectively). 
The calculations are done for both types of wavefunction Ans\"atze: UCCSD and heuristic.
In the UCCSD case, the circuit corresponding to the exponentiation of the operators $\hat{T}_1$ and $\hat{T}_2$ (in Eqs.~\eqref{eq:t1} and~\eqref{eq:t2}) is the same for both Hamiltonians and therefore we do not expect any benefit from the p/h approach in terms of the reduction of the gate count. 
However, the optimization of the parameters becomes notably more efficient using the p/h Hamiltonian.
The number of BFGS iterations required to achieve a convergence of $10^{-7}$~Ha decreases from 53 for the plain Hamiltonian to 27 in the p/h case.
Most importantly, the number of circuit measurements required for the full optimization also drops from $3383 \times N_s$ for the plain Hamiltonian to only $1471\times N_s$ in the p/h formalism, where $N_s$ is the number of shots used to evaluate the energy.
Combining these effects, we observe an overall speed-up in the p/h implementation of UCCSD of about a factor 2 to 3.

\begin{table} [h] 
{\def\arraystretch{1.15}\tabcolsep=2pt
    \caption{Statistics on the simulation of the ground state energy for $\rm{H_2}$ using the original Hamiltonian in second quantization (Eq.~\eqref{Eq:H_sec_quant}) and the p/h Hamiltonian (Eq.~\eqref{eq:HF_H_ph}). Results are given for the UCCSD expansion (with a single Trotter step, see Section~\ref{subsec:Trotter}) and three heuristic approaches using ${U}^{(1)}_{\rm ent}$, ${U}^{(2)}_{\rm ent}$ and ${U}^{(3)}_{\rm ent}$ gates. Comparison is based on: number of Pauli strings evaluations for determination of the gradients in parameter space (Eval.), number of optimization steps in the BFGS algorithm (Iter.), number of single-qubit (1qG) and two-qubit (2qG) gates, total number of parameters (Par.), and the number of entangling blocks (D).\\}
        \begin{tabular}{c|c|cc|cc|cc}
        \hline
        \multicolumn{8}{c}{\it{SQ Hamiltonian}} \\ \hline \hline
        \multicolumn{1}{c}{\ } & \multicolumn{1}{c}{UCCSD} &  \multicolumn{6}{c}{Heuristic} \\ \hline
        \multicolumn{1}{c}{\ } & \multicolumn{1}{c}{\ } &  \multicolumn{2}{c}{${U}^{(1)}_{\rm ent}$} & \multicolumn{2}{c}{${U}^{(2)}_{\rm ent}$} & \multicolumn{2}{c}{${U}^{(3)}_{\rm ent}$}\\\hline
        {}       & Full \ & Equil \  & Full \ & Equil \ & Full \ & Equil \ & Full \ \\
        Eval.($10^3$)    & 3.3    & 35 & 38.4 & 12.5  & 19    & 28.6     & -\\
        Iter.    &  53      & 740   & 812 &  59    & 88    & 420     & -\\
        1qG &  708     & 0     & 0  &  64    & 96   & 112     & >144\\
        2qG &  608     & 56    & 70 &  56    & 84   & 392     & >504\\
        Par.   &  15   & 112   & 140 & 120   & 180  & 112     & >144\\ 
        D   & - & 8   &  10 & 8   & 12  & 14     & >18\\ \hline
        \multicolumn{8}{c}{ \ }  \\ \hline
        \multicolumn{8}{c}{\it{SQ Particle-Hole Hamiltonian}} \\ \hline \hline 
        \multicolumn{1}{c}{\ } & \multicolumn{1}{c}{UCCSD} &  \multicolumn{6}{c}{Heuristic} \\ \hline
        \multicolumn{1}{c}{\ } & \multicolumn{1}{c}{\ }  &  \multicolumn{2}{c}{${U}^{(1)}_{\rm ent}$} & \multicolumn{2}{c}{${U}^{(2)}_{\rm ent}$} & \multicolumn{2}{c}{${U}^{(3)}_{\rm ent}$}\\
        \hline
        {}       & Full \ & Equil \  & Full \ & Equil \ & Full \ & Equil \ & Full\\
        Eval.($10^3$)    & 1.5    & 21.4 & 32.1 & 10.7  & 14.8    & 17.2       & -\\
        Iter.    &  27      & 421   & 578 & 52     & 78   & 254    & -\\
        1qG & 708      & 0     & 0 & 64     & 96   & 112     & >144\\
        2qG & 608      & 42    & 70 & 56     & 84   & 392      & >504\\
        Par.   &  15      & 84    & 140 & 120    & 180   & 112      & >144\\
        D   & -     & 6  & 10 & 8   & 12  & 14     & >18\\ \hline
        \end{tabular}
    \label{table:Hamiltonian-comparison}
}
\end{table}

The situation is similar in the heuristic wavefunction approach using either ${U}^{(1)}_{\rm ent}$ or ${U}^{(2)}_{\rm ent}$ entangler blocks. 
In these cases, the number of entangler blocks, $D$ in Eq.~\eqref{eq:entangler}, is increased until convergence of the final energy is reached, i.e., $|E_{\rm heur}^{D}-E_{\text{exact}}|< \epsilon$, where $E_{\rm heur}^{D}$ is the energy of the heuristic approach with $D$ blocks and $\epsilon$ is the chemical accuracy.
Column `Equil' in Table~\ref{table:Hamiltonian-comparison} reports the values required for convergence at the equilibrium position ($\sim 0.7$~\AA), while `Full' refers to the numbers obtained when convergence is imposed 
over the entire dissociation path (maximum value over the entire dissociation path).
At each value of $D$, the number of iterations of the classical optimizer (`Iters' in Table~\ref{table:Hamiltonian-comparison}) is determined by the convergence criteria set for the energy difference between two consecutive iterations ($< 10^{-7}$ Ha).
We first observe that using the p/h Hamiltonian (at the equilibrium distance, `Equil.') the same accuracy ($10^{-7}$~Ha) can be achieved with only 6 entangler blocks compared to the 8 required when using the plain Hamiltonian.
This has the effect of reducing, at least in the case of the ${U}_{\rm{ent}}^{(1)}$, both the number of parameters (from 112 to 84) and the total number of gate operations (from 56 to 42).
As in the UCCSD case, also in the heuristic approach the number of iterations as well as the number of circuit evaluations drop significantly when using the p/h Hamiltonian. 
For the case in which chemical accuracy is required at all distances 
(Table~\ref{table:Hamiltonian-comparison}, columns `Full'), we see an overall gain for the heuristic implementation of the p/h approach of about a factor 3 to 4 compared to standard SQ Hamiltonian (Eq.~\eqref{Eq:H_sec_quant}).

We also note that the convergence with the CNOT entanglers requires in general a larger number of resources and in some cases (CNOT/Full in Table~\ref{table:Hamiltonian-comparison}) it is not possible to reach a solution within chemical accuracy with less than 18 blocks. 

\subsection{The UCCSD Ansatz}

The quantum algorithm based on the UCCSD Ansatz for the molecular wavefunction is obtained by directly mapping the exponentials in Eqs.~\eqref{eq:t1} and~\eqref{eq:t2} into the corresponding quantum circuits (see Fig.~\ref{fig:UCCSD_circ_T1}).
In this work we investigate two developments of the UCC approach that lead to a simplification of the corresponding quantum algorithm by reducing the circuit depth.
The first one is based on the restriction of the Hilbert space using an active space as described in Section~\ref{subsec:Active_space_UCCSD}. 
The other one consists on the exploitation of the variational character of the UCCSD Anzatz, which introduces the possibility of `absorbing' the effect of some approximations (e.g., the Trotter decomposition)  through a suitable modification of parameters used to span the wavefunction space.

\subsubsection{Active space in the UCCSD approach}
\label{subsec:Active_space_UCCSD}

We start with the simulation of the hydrogen molecule. As mentioned  above, the 6-31G basis set~\cite{Ditchfield1971} leads to a Hilbert space spanned by 8 HF orbitals, 2 occupied and 6 virtuals (i.e., unoccupied). 
Note that in order to keep a one-to-one correspondence with the classical UCCSD algorithm we do not perform any additional reduction of the Hamiltonian as was done in previous studies~\cite{moll_optimizing_2016}.

\begin{figure}[h]
    \centering
    \includegraphics[width = \columnwidth]{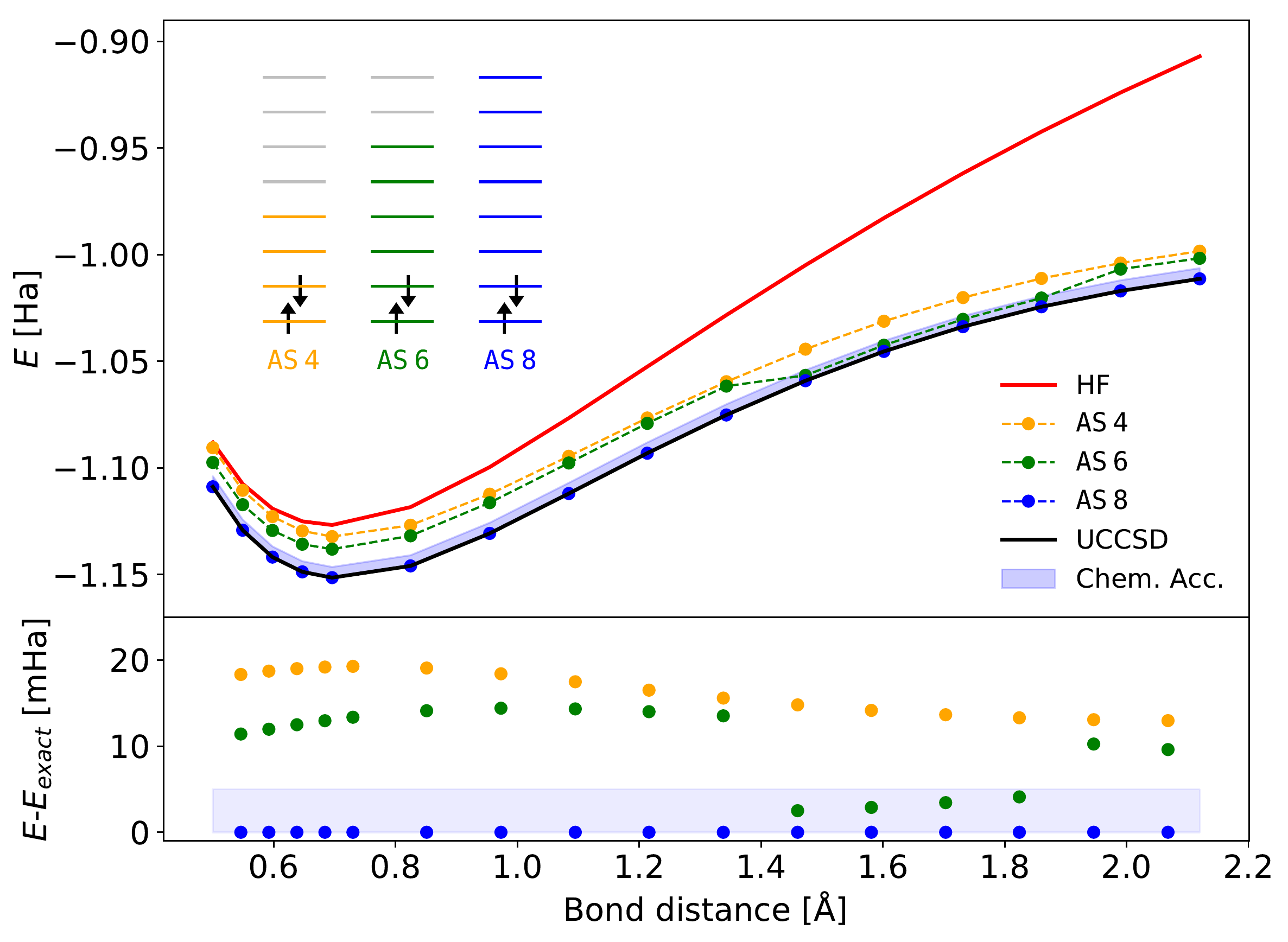}
    \caption {Upper panel: Dissociation profile of the $\rm{H_2}$ molecule for different definitions of the active space (AS).
    AS 4 (orange): only 2 occupied and 2 virtual orbitals are considered in the definition of the $\hat T_1$ and $\hat T_2$ operators; 
    AS 6 (green): 2 occupied and 4 virtual orbitals;
    AS 8 (blue): 2 occupied and 6 virtual orbitals.
    The red curve corresponds to the reference HF calculation and the black one is the analytic solution evaluated using the p/h Hamiltonian expanded in the full (12 qubit) space.
    Lower panel: Corresponding energy errors along the dissociation profile. 
    The blue shaded area corresponds to the energy range within chemical accuracy.}
    \label{fig:UCCSD_AS_H2}
\end{figure}

Fig.~\ref{fig:UCCSD_AS_H2} shows the dissociation profile for the ${\rm H_2}$ molecule computed using VQE approach and the UCCSD Ansatz with different sizes of the active space (AS) ranging from a minimum of 4 to the full space.
For all choices of the active space, we observe a noticeable improvement of the evaluated dissociation curve compared to the zero-order approximation given by HF profile (red line).
More interestingly, the results obtained with the smallest AS (AS4) already provide a qualitatively correct curve that runs approximately in parallel to the exact solution obtained by diagonalizing the p/h Hamiltonian in the chosen basis set (6-31G).
By increasing the size of the active space we observe a gradual improvement of the quality of computed dissociation curve with errors that approaches chemical accuracy (set to $0.5 \times 10^{-2}$~Ha).

\begin{figure}[h]
    \centering
    \includegraphics[width = \columnwidth]{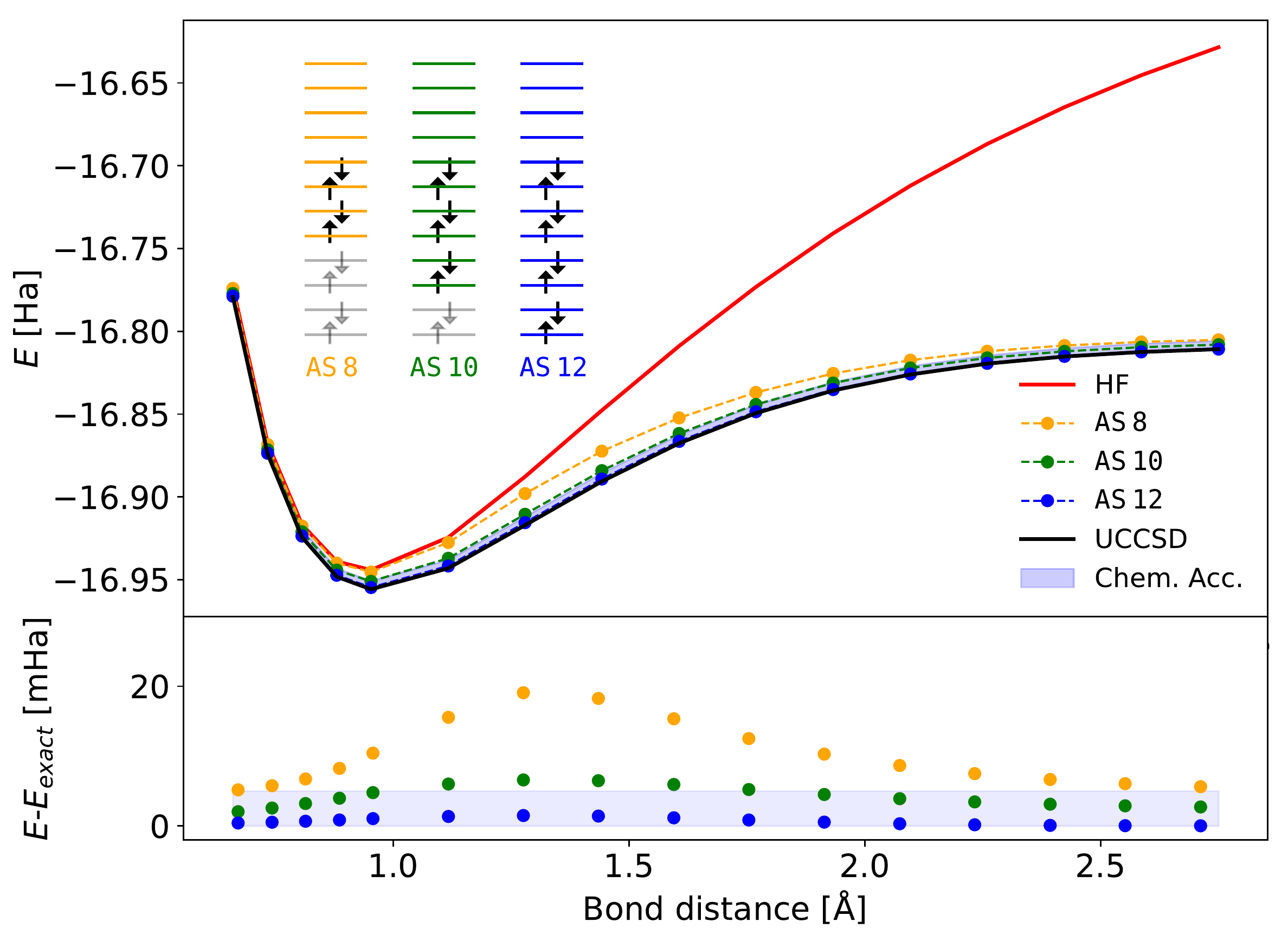}
    \caption{Upper panel: Dissociation profile of the $\rm{H_2O}$ molecule for different definitions of the active space (AS).
    AS 8 (orange): 4 HF orbitals (starting form the highest occupied one, see inset) and all virtual orbitals are considered in the definition of the $\hat T_1$ and $\hat T_2$ operators; 
    AS 10 (green): 6 occupied and all virtual orbitals;
    AS 12 (blue): 8 occupied and all virtual orbitals.
    The red curve corresponds to the reference HF calculation and the black one is the analytic solution evaluated using the p/h Hamiltonian expanded in the full (12 qubit) space.
    Lower panel: Corresponding energy errors along the dissociation profile. 
    The blue shaded area corresponds to the energy range within chemical accuracy.}
    \label{fig:UCCSD_AS_H2O}
\end{figure}

Fig.~\ref{fig:UCCSD_AS_H2O} reports the same energy profiles for the asymmetric stretch of a O-H bond of a water molecule.
The exact solution is computed using a direct diagonalization of the p/h Hamiltonian generated using the first 12 lowest energy HF orbitals, among which 8 are occupied. 
In this case the different active spaces (AS4, AS6, AS8) are defined by the size of occupied HF orbitals included in the expansions for the $\hat T_1$ and $\hat T_2$ operators (see inset of Fig.~\ref{fig:UCCSD_AS_H2O}).
As for the case of the hydrogen molecule, the correction to the HF profile is large for all choices of the active space and the error decreases monotonically as the number of electrons included increases.
The largest deviations are measured for the intermediate bond lengths, while the error goes below the line delimiting chemical accuracy (shaded blue region) at the distances below the equilibrium value ($< 1$~\AA) and in the dissociation limit ($> 2$~\AA).
The intermediate range corresponds to geometries close to the so-called Coulson-Fisher point where spin-symmetry breaking can occur~\cite{Gunnarsson1976}.

\subsubsection{Trotter error in UCCSD}
\label{subsec:Trotter}

A major drawback of the UCCSD implementation are the errors introduced by the Trotter factorization of Eq.~\eqref{eq:UCCmain}.

In this section, we investigate the convergence of the energy $E_{\rm UCCSD}^{\text{an}/n}$ as a function of the number of Trotter steps, $n$. 
The expansion in Eq.~\eqref{eq:UCCmain} can be written 
\begin{equation}
    e^{(\hat T_1- \hat T_1^\dagger) + (\hat T_2- \hat T_2^\dagger)}
    \approx 
    \left( 
    e^{(\hat T_1-\hat T_1^\dagger)/n} \, e^{(\hat T_2-\hat T_2^\dagger)/n} 
    \right)^n \, ,
    \label{eq:Trotter_approx}
\end{equation}
and becomes exact in the limit $n\rightarrow \infty$.
To this end, we designed a series of test calculations, which probe the origin of the different errors arising from the truncation at second order in UCCSD and the used of the Trotter decomposition.

To simplify the discussion, we report a summary of the different simulations and the corresponding approximations in Table~\ref{table:UCCSD_simu}.
\begin{table} [h] 
{\def\arraystretch{1.4}\tabcolsep=2.pt
    \caption{Summary of the different simulations used to describe the approximations in UCCSD. More details in the text.}
    \begin{tabular}{c| c| c }
    \hline
    \multicolumn{1}{l}{ } & \multicolumn{1}{c}{Description} & \multicolumn{1}{c}{Approximations} \\
    \hline
    \hline
    $  E_{\rm diag}$ & diagonalization of $\hat{H}^{p/h}$ & Exact \\
    \hline
    \multirow{ 2}{*}{$ {E}_{\rm UCCSD}^{\rm an}$} & analytic UCCSD (matrix repr.)  & \multirow{ 2}{*}{Truncation at $\hat T_2$} \\ 
    & using exact exponentiation  & \\
    \hline
    \multirow{ 2}{*}{${E}_{\rm UCCSD}^{{\rm an}/n}$} & analytic UCCSD (matrix repr.)   &  Truncation at $\hat T_2$ \\
    & using $n$ Trotter steps  & \& Trotter error\\
    \hline
    \multirow{ 2}{*}{${E}_{\rm UCCSD}^{{\rm circ}/n}$} & UCCSD/VQE in circuits  &  Truncation at $\hat T_2$ \\ 
    &   using $n$ Trotter steps  & \& Trotter error \\
    \hline
    \end{tabular}
    \label{table:UCCSD_simu}
}
\end{table}
As a reference, we take the first eigenvalue from the diagonalization of the p/h Hamiltonian in Eq.~\eqref{eq:HF_H_ph}, $E_{\rm diag}$. 
The energy $E_{\rm UCCSD}^{\rm an}$ is evaluated using optimization of the UCCSD coefficients in the matrix representation of the expansion in Eqs.~\eqref{eq:t1} and~\eqref{eq:t2} (exact exponentiation in Eq.~\eqref{eq:UCCmain}). 
The difference $E_{\rm diag}$- $E_{\rm UCCSD}^{\rm an}$ is a measure for the error introduced by the truncation of the excitation operator at second order $\hat T=\hat T_1+\hat T_2$.
Finally, the energies $E_{\rm UCCSD}^{{\rm an}/n}$ and $E_{\rm UCCSD}^{{\rm circ}/n}$  are computed using the approximated Trotter expansion with $n$ steps, Eq.~\eqref{eq:Trotter_approx}, in the matrix and circuit representations (Fig.~\ref{fig:UCCSD_circ_T1}), respectively.
This energy difference 
provides a clear measure of the error introduced by the truncation of the Trotter expansion to $n$-order. 
Due to the perfect agreement between the values of $E_{\rm UCCSD}^{{\rm an}/n}$ and $E_{\rm UCCSD}^{{\rm circ}/n}$ (data not shown), in the following we will only use $E_{\rm UCCSD}^{{\rm circ}/n}$.

\begin{figure}[h]
    \centering
    \includegraphics[width = \columnwidth]{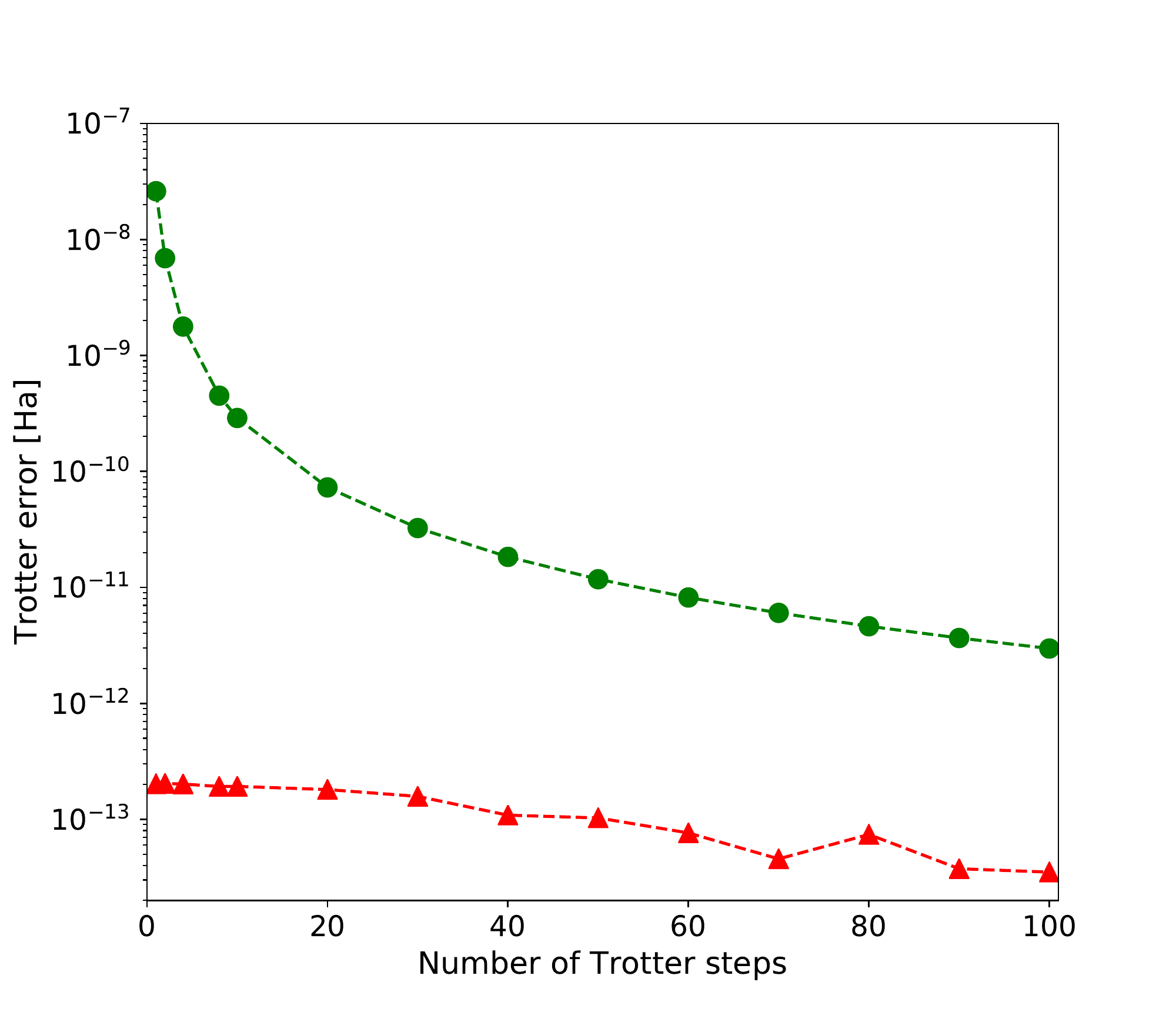}
    \caption{
    Convergence of the Trotter error as a function of the Trotter expansion coefficient $n$ in Eq.~\eqref{eq:Trotter_approx}
    for the UCCSD energy of $\rm H_2$ at a bond length of $0.592$~\AA.
    The reference energy, $E_{\text{exact}}$, corresponds to $E_{\rm diag}$ from Table~\ref{table:UCCSD_simu}. Green circles: analytic dependence of the Trotter error ($E_{\rm UCCSD}^{opt/n}$ in Eq.~\eqref{eq:E_optn_UCCSD}). Red triangles: Trotter errors obtained after the optimization of the angles $\vec{\theta}$ at each value of $n$ using the VQE approach  ($E_{\rm UCCSD}^{{\rm circ}/n}(n)$ in Eq.~\eqref{eq:E_circn_UCCSD}).}
    \label{fig:trotter_error}
\end{figure}

In Fig.~\ref{fig:trotter_error}, we summarize the results for the Trotter approximation in the Hydrogen molecule.
Without loss of generality, we will focus on a single geometry corresponding to a bond length of 0.592~\AA.
As reference energy we take the analytic value $E_{\rm diag}$. 
The exact UCCSD expansion coefficients are obtained using the VQE algorithm and the analytic representation of the exponentiations in Eqs.~\eqref{eq:t1} and~\eqref{eq:t2}. 
Using these coefficients $\theta_{opt}$
(corresponding to the energy ${E}_{\rm UCCSD}^{\rm an}$), we recompute the energies 
\begin{equation}
 E^{opt/n}_{\rm UCCSD} (n ) = \langle \psi_{Tr} (\vec \theta_{opt}, n )  \vert \hat{H}^{p/h}  \vert \psi_{Tr} (\vec \theta_{opt}, n)  \rangle 
 \label{eq:E_optn_UCCSD}
\end{equation}
using a $n$-steps Trotter expansion. 
$\vert \psi_{Tr}(\vec \theta_{opt}, n) \rangle$ corresponds to the state obtained using the the optimized angles $\vec \theta_{opt}$ and the right-hand side of Eq.~\eqref{eq:Trotter_approx} (with variable $n$) instead of the exact expression (left-hand side of Eq.~\eqref{eq:Trotter_approx}).
The error with respect to the exact energy is given by the green circles in Fig.~\ref{fig:trotter_error}.
Interestingly enough, when we apply the VQE approach and therefore a full reoptimization of the parameters at each value of $n$, we obtain the energies (Fig.~\ref{fig:trotter_error}, red triangles)
\begin{equation}
{E}_{\rm UCCSD}^{{\rm circ}/n} (n) = \min\limits_{\vec\theta} \langle \psi_{Tr} (\vec \theta, n ) \vert \hat{H}^{p/h} \vert \psi_{Tr} (\vec \theta, n)  \rangle \, ,
 \label{eq:E_circn_UCCSD}
\end{equation}
which show
a negligibly small error $|E_{\rm diag}$-$E_{\rm UCCSD}^{{\rm circ}/n}(n)|$, independent from the number of Trotter steps  (see also~\cite{Wecker2015a}).
Already with $n=1$ we measure an error smaller than $10^{-10}$~Ha, i.e., well below chemical accuracy.
This result implies that the flexibility introduced by the large number of parameters in VQE can variationally absorb the Trotter error even at $n=1$.
Therefore, the UCCSD approach based on VQE algorithm deviates from what would be the one-to-one mapping of the classical CCSD expansion, becoming a mixed CLA-HEA approach that we name q-UCCSD.
This very important result can enormously impact  the future implementation of the UCCSD Ansatz in quantum circuits for large molecules since the Trotter expansion can be implemented in one step, reducing the circuit depth.

\subsection{The heuristic Ansatz}

\begin{figure}[h]
    \centering
    \includegraphics[width = \columnwidth]{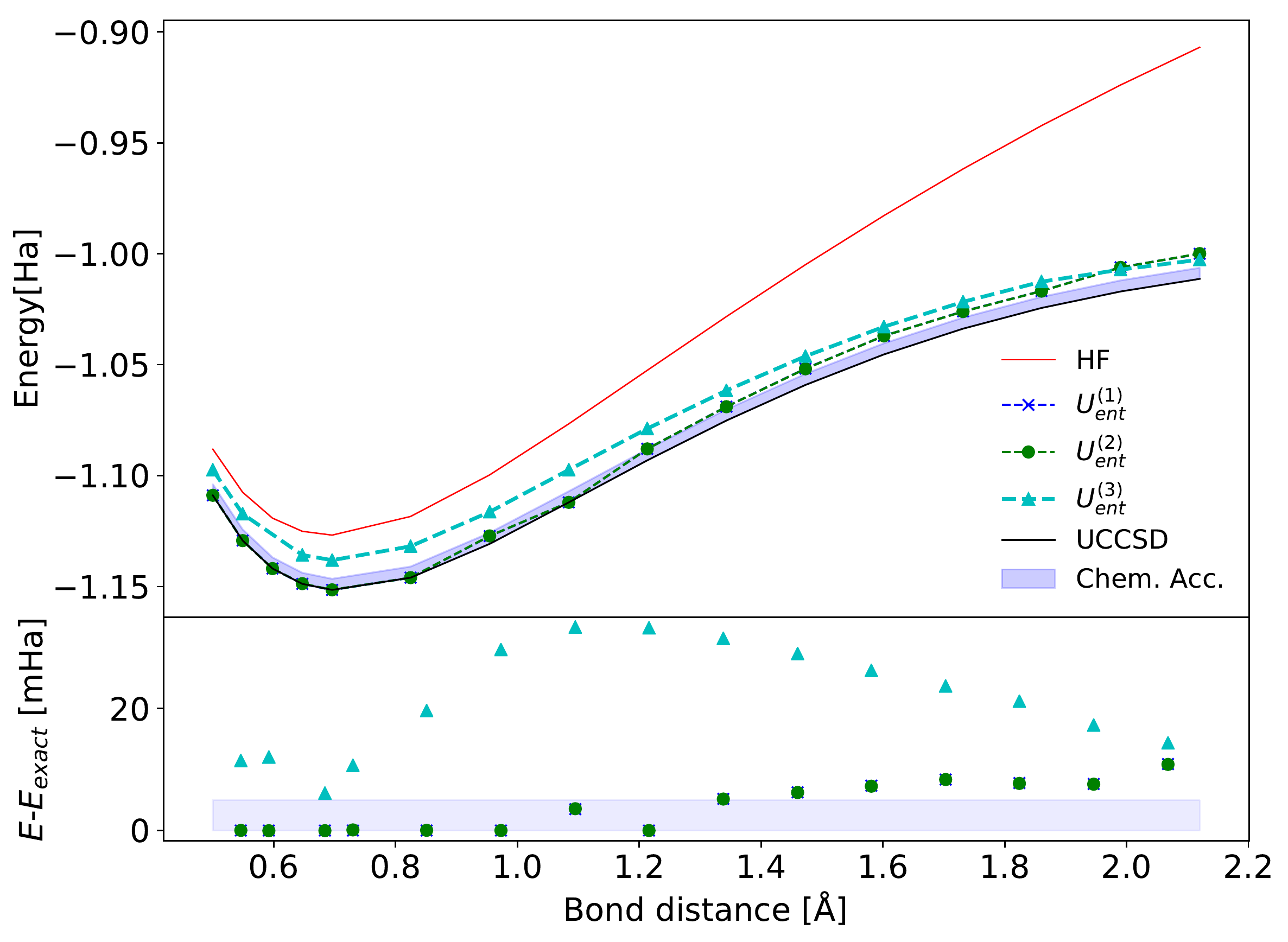}
     \caption{Dissociation profiles of the $\rm{H_2}$ molecule computed using the p/h Hamiltonian and the heuristic Ansatz for the trial wavefunction. 
     The blue crosses and green dots are obtained using the particle-conserving entanglers $ U^{(1)}_{\rm ent}$ and $ U^{(2)}_{\rm ent}$, respectively. The cyan triangles are computed using the non particle-conserving
     operator circuit $ U^{(3)}_{\rm ent}$.
     The shaded area corresponds to the chemical accuracy energy range.
     For all entrangler types, the number of repeated blocks, $D$ in Eq.~\eqref{eq:entangler}, was set to 8.
     }
    \label{fig:H2_profile_heuristic}
\end{figure}

\begin{figure}[h]
    \centering
    \includegraphics[width = \columnwidth]{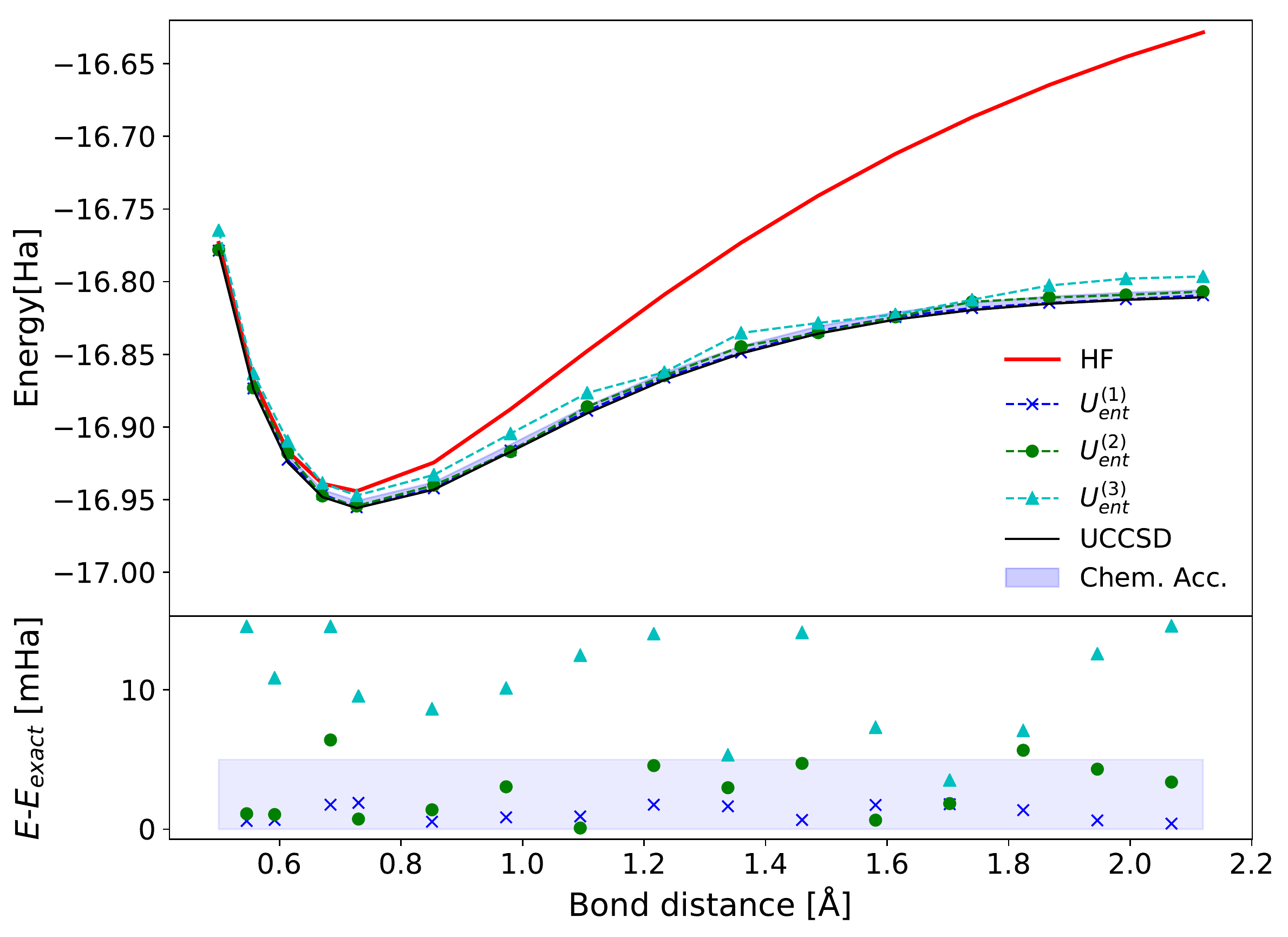}
    \caption{Dissociation profiles of the $\rm{H_2O}$ molecule computed using the p/h Hamiltonian and the heuristic Ansatz for the trial wavefunction. The blue crosses and green dots are obtained using the particle-conserving entanglers $ U^{(1)}_{\rm ent}$ and $ U^{(2)}_{\rm ent}$, respectively. The cyan triangles are computed using the non particle-conserving
     operator circuit $ U^{(3)}_{\rm ent}$. The shaded area corresponds to the chemical accuracy energy range. 
     For all entrangler types, the number of repeated blocks, $D$ in Eq.~\eqref{eq:entangler}, was set to 20.
     }
    \label{fig:H2O_profile_heuristic}
\end{figure}

In this Section we further develop methods of the heuristic wavefunction Ansatz that were introduced in~\cite{kandala_hardware-efficient_2017}.
As in the case of UCCSD, we begin by encoding the qubits the occupancy of the molecular HF orbitals instead of atomic ones.
Then, following the developments in the theory section, we combined the heuristic VQE approach with the p/h Hamiltonian, which provides a more efficient starting point for optimization of the ground state energy.
Within this framework, we also made use of the ECPs to decrease the number of degrees of freedom.
In particular, we compared the level of accuracy and the efficiency of the three entangler blocks defined in Section~\ref{sss:Heuristic} (see Fig.~\ref{fig:circuits}).
In the case of the non-particle conserving entangler (${U}_{\rm CNOT}$) the chemical potential term of Eq.~\eqref{eq:chem_pot} is added to the p/h Hamiltonian.
The number of entangler blocks for each heuristic Ansatz is fixed by imposing an energy accuracy of $10^{-7}$ Ha at the equilibrium position. By increasing the number of blocks it is possible to achieve convergence within chemical accuracy along the entire dissociation profile, at the cost of further increasing the circuit depth (see Table~\ref{table:Hamiltonian-comparison}).
The dissociation curve for the $\rm{H_2}$ and $\rm{H_2O}$ molecules computed using the p/h Hamiltonian and the three entangler blocks $U_{\rm ent}^{(1-3)}$ are shown respectively in Figs.~\ref{fig:H2_profile_heuristic} and ~\ref{fig:H2O_profile_heuristic}.
As reference, we also plot the HF profile and the reference curve obtained from the diagonalization of the p/h Hamiltonian. 
We observe that both particle-conserving entanglers ($ U^{(1)}_{\rm ent}$, and $ U^{(2)}_{\rm ent}$) produce very similar profiles with very small deviations around the equilibrium position that become increasingly larger as the distance between the two hydrogen atoms increases. 
The non particle-conserving entangler ($ U^{(3)}_{\rm ent}$) shows larger deviations over the entire distance range (compared to the exact energy, $E_{\rm diag}$). 
The reason for this discrepancy can be twofold. 
It may arise form a sampling deficiency of the CNOT gate sequence, or by an intrinsic convergence problem related to the much larger size of the sampling space (the Fock space with variable number of electrons).
Comparing with the UCCSD results, we observe that with the quoted number of entangler blocks we do not achieve chemical accuracy at large distances ($R>1.3$~\AA). On the other hand, in the case of the water molecule both $\hat U^{(1)}_{\rm ent}$ and $\hat U^{(2)}_{\rm ent}$ entanglers are within chemical accuracy all along the dissociation path. 

\section{CONCLUSIONS}
\label{section_conclusions}
In this work we examine the implementation of different quantum algorithms for the calculation of the ground state energy of simple molecular systems in quantum computers. 
In particular, we show that the reformulation of the molecular Hamiltonian in second quantization using the particle-hole (p/h) picture brings important advantages in terms of computational efficiency and accuracy.
 By shifting the reference state from the vacuum to the HF wavefunction, this approach
provides a better starting point for a systematic expansion of the molecular wavefunction, which leads to a faster convergence of the correlation energy  $E_{\text{corr}}=E_{\text{GS}}-E_{\text{HF}}$.
We also analyzed the effects of restricting the Hilbert space to the subspace of chemically active valence electrons.
By replacing core electrons with the corresponding effective core potentials, we obtain a modified p/h-Hamiltonian, which can reproduce ground state molecular energies within chemical accuracy using a reduced number of qubits.

Additionally, we also investigate the implementation of two different wavefunction Ans\"atze for the optimization of the
ground state energy within the Variational Quantum Eigensolver (VQE) algorithm.
The first one is based on  
an adaptation of 
the Coupled Cluster expansion series used in quantum chemistry. 
We show that within the VQE framework the exponentiation of the cluster operators (see Eq.~\eqref{eq:UCCmain}) can be efficiently approximated with a single Trotter step, while keeping 
a good level of accuracy
for the ground state energy (errors below $10^{-10}$ Ha in simulations). 
This surprising result can be explained with the flexibility of the variational quantum algorithm, which relies on an efficient parametrization of the Hilbert space that can `absorb' the error introduced by the Trotter approximation.
As such, this VQE implementation of the CC approach looses its original one-to-one correspondence with the original classical algorithm and acquires a different, variational, character. 
For this reason we introduced the q-UCC acronym to define the quantum version of the classical CC approach.
The use of a single Trotter step has also important implications for the implementation of the CC algorithm in real hardware since it implies a drastic reduction (of about a factor $10^3$) in the number of gates required.
The second approach is based on the heuristic wavefunction expansion introduced originally in Ref.~\cite{kandala_hardware-efficient_2017}.
In this case, we proposed a set of specialized exchange-type two-qubit gates, which substantially improve  the efficiency of the entangler blocks used to generate the trial wavefunctions in the VQE approach.
The success of both exchange-type gates is related to their particle-conserving property, which enables to restrict the sampling of the Hilbert space within the relevant subspace with the correct number of electrons.

We apply these developments to the study of the dissociation curves of molecular hydrogen ($\rm H_2$) and water ($\rm H_2O$). 
The p/h Hamiltonian shows clear advantages compared to original Hamiltonian in terms of implementation (shorter circuit depth) and convergence efficiency (smaller number of iterations).
We showed that both wavefunction Ans\"atze can converge the ground state energy within chemical accuracy. In the q-UCC approach, the possibility to define active spaces, which confine the excitations to a subset of the occupied and virtual orbitals, can be used to further reduce the gate count while keeping a good and controllable level of accuracy.

In conclusion, we show that within the p/h-formalism it is possible to design valid quantum algorithms for the solution of electronic structure problems in near-term quantum computers with a favourable scaling in terms of parameters and gate counts.

\section{ACKNOWLEDGEMENTS}
The authors thank Stefan Woerner, Pauline Ollitrault, Walter Riess, Peter Mueller, Andreas Woitzik, Filip Wudarski, Andreas Buchleitner, Marco Pistoia, Abhinav Kandala, Julia Rice, Stephen Wood and Jay Gambetta for useful discussions.

\bibliography{QC_Chem_Paperbib_new.bib}

\begin{thebibliography}{63}%
\makeatletter
\providecommand \@ifxundefined [1]{%
 \@ifx{#1\undefined}
}%
\providecommand \@ifnum [1]{%
 \ifnum #1\expandafter \@firstoftwo
 \else \expandafter \@secondoftwo
 \fi
}%
\providecommand \@ifx [1]{%
 \ifx #1\expandafter \@firstoftwo
 \else \expandafter \@secondoftwo
 \fi
}%
\providecommand \natexlab [1]{#1}%
\providecommand \enquote  [1]{``#1''}%
\providecommand \bibnamefont  [1]{#1}%
\providecommand \bibfnamefont [1]{#1}%
\providecommand \citenamefont [1]{#1}%
\providecommand \href@noop [0]{\@secondoftwo}%
\providecommand \href [0]{\begingroup \@sanitize@url \@href}%
\providecommand \@href[1]{\@@startlink{#1}\@@href}%
\providecommand \@@href[1]{\endgroup#1\@@endlink}%
\providecommand \@sanitize@url [0]{\catcode `\\12\catcode `\$12\catcode
  `\&12\catcode `\#12\catcode `\^12\catcode `\_12\catcode `\%12\relax}%
\providecommand \@@startlink[1]{}%
\providecommand \@@endlink[0]{}%
\providecommand \url  [0]{\begingroup\@sanitize@url \@url }%
\providecommand \@url [1]{\endgroup\@href {#1}{\urlprefix }}%
\providecommand \urlprefix  [0]{URL }%
\providecommand \Eprint [0]{\href }%
\providecommand \doibase [0]{http://dx.doi.org/}%
\providecommand \selectlanguage [0]{\@gobble}%
\providecommand \bibinfo  [0]{\@secondoftwo}%
\providecommand \bibfield  [0]{\@secondoftwo}%
\providecommand \translation [1]{[#1]}%
\providecommand \BibitemOpen [0]{}%
\providecommand \bibitemStop [0]{}%
\providecommand \bibitemNoStop [0]{.\EOS\space}%
\providecommand \EOS [0]{\spacefactor3000\relax}%
\providecommand \BibitemShut  [1]{\csname bibitem#1\endcsname}%
\let\auto@bib@innerbib\@empty
\bibitem [{\citenamefont {Lloyd}(1996)}]{lloyd_universal_1996}%
  \BibitemOpen
  \bibfield  {author} {\bibinfo {author} {\bibfnamefont {S.}~\bibnamefont
  {Lloyd}},\ }\href {\doibase 10.1126/science.273.5278.1073} {\bibfield
  {journal} {\bibinfo  {journal} {Science}\ }\textbf {\bibinfo {volume}
  {273}},\ \bibinfo {pages} {1073} (\bibinfo {year} {1996})}\BibitemShut
  {NoStop}%
\bibitem [{\citenamefont {Preskill}(2018)}]{preskill18}%
  \BibitemOpen
  \bibfield  {author} {\bibinfo {author} {\bibfnamefont {J.}~\bibnamefont
  {Preskill}},\ }\href {http://arxiv.org/pdf/1801.00862v1} {\bibfield
  {journal} {\bibinfo  {journal} {arXiv:1801.00862v1[quant-ph]}\ } (\bibinfo
  {year} {2018})}\BibitemShut {NoStop}%
\bibitem [{\citenamefont {Nielsen}\ and\ \citenamefont
  {Chuang}(2011)}]{Nielse_Chuang2011}%
  \BibitemOpen
  \bibfield  {author} {\bibinfo {author} {\bibfnamefont {M.~A.}\ \bibnamefont
  {Nielsen}}\ and\ \bibinfo {author} {\bibfnamefont {I.~L.}\ \bibnamefont
  {Chuang}},\ }\href@noop {} {\emph {\bibinfo {title} {Quantum Computation and
  Quantum Information: 10th Anniversary Edition}}},\ \bibinfo {edition} {10th}\
  ed.\ (\bibinfo  {publisher} {Cambridge University Press},\ \bibinfo {address}
  {New York, NY, USA},\ \bibinfo {year} {2011})\BibitemShut {NoStop}%
\bibitem [{\citenamefont {Abrams}\ and\ \citenamefont
  {Lloyd}(1999)}]{abrams_quantum_1999}%
  \BibitemOpen
  \bibfield  {author} {\bibinfo {author} {\bibfnamefont {D.~S.}\ \bibnamefont
  {Abrams}}\ and\ \bibinfo {author} {\bibfnamefont {S.}~\bibnamefont {Lloyd}},\
  }\href {\doibase 10.1103/PhysRevLett.83.5162} {\bibfield  {journal} {\bibinfo
   {journal} {Phys. Rev. Lett.}\ }\textbf {\bibinfo {volume} {83}},\ \bibinfo
  {pages} {5162} (\bibinfo {year} {1999})}\BibitemShut {NoStop}%
\bibitem [{\citenamefont {Brassard}\ \emph {et~al.}(2000)\citenamefont
  {Brassard}, \citenamefont {Hoyer}, \citenamefont {Mosca},\ and\ \citenamefont
  {Tapp}}]{Brassard2000}%
  \BibitemOpen
  \bibfield  {author} {\bibinfo {author} {\bibfnamefont {G.}~\bibnamefont
  {Brassard}}, \bibinfo {author} {\bibfnamefont {P.}~\bibnamefont {Hoyer}},
  \bibinfo {author} {\bibfnamefont {M.}~\bibnamefont {Mosca}}, \ and\ \bibinfo
  {author} {\bibfnamefont {A.}~\bibnamefont {Tapp}},\ }\href
  {https://arxiv.org/abs/quant-ph/0005055} {\bibfield  {journal} {\bibinfo
  {journal} {arXiv:quant-ph/0005055}\ } (\bibinfo {year} {2000})}\BibitemShut
  {NoStop}%
\bibitem [{\citenamefont {H\o{}yer}(2000)}]{Hoyer2000}%
  \BibitemOpen
  \bibfield  {author} {\bibinfo {author} {\bibfnamefont {P.}~\bibnamefont
  {H\o{}yer}},\ }\href {\doibase 10.1103/PhysRevA.62.052304} {\bibfield
  {journal} {\bibinfo  {journal} {Phys. Rev. A}\ }\textbf {\bibinfo {volume}
  {62}},\ \bibinfo {pages} {052304} (\bibinfo {year} {2000})}\BibitemShut
  {NoStop}%
\bibitem [{\citenamefont {Abrams}\ and\ \citenamefont
  {Lloyd}(1997)}]{abrams_simulation_1997}%
  \BibitemOpen
  \bibfield  {author} {\bibinfo {author} {\bibfnamefont {D.~S.}\ \bibnamefont
  {Abrams}}\ and\ \bibinfo {author} {\bibfnamefont {S.}~\bibnamefont {Lloyd}},\
  }\href {\doibase 10.1103/PhysRevLett.79.2586} {\bibfield  {journal} {\bibinfo
   {journal} {Phys. Rev. Lett.}\ }\textbf {\bibinfo {volume} {79}},\ \bibinfo
  {pages} {2586} (\bibinfo {year} {1997})}\BibitemShut {NoStop}%
\bibitem [{\citenamefont {Lanyon}\ \emph {et~al.}(2010)\citenamefont {Lanyon},
  \citenamefont {Whitfield}, \citenamefont {Gillett}, \citenamefont {Goggin},
  \citenamefont {Almeida}, \citenamefont {Kassal}, \citenamefont {Biamonte},
  \citenamefont {Mohseni}, \citenamefont {Powell}, \citenamefont {Barbieri},
  \citenamefont {Aspuru-Guzik},\ and\ \citenamefont
  {White}}]{lanyon_towards_2010}%
  \BibitemOpen
  \bibfield  {author} {\bibinfo {author} {\bibfnamefont {B.~P.}\ \bibnamefont
  {Lanyon}}, \bibinfo {author} {\bibfnamefont {J.~D.}\ \bibnamefont
  {Whitfield}}, \bibinfo {author} {\bibfnamefont {G.~G.}\ \bibnamefont
  {Gillett}}, \bibinfo {author} {\bibfnamefont {M.~E.}\ \bibnamefont {Goggin}},
  \bibinfo {author} {\bibfnamefont {M.~P.}\ \bibnamefont {Almeida}}, \bibinfo
  {author} {\bibfnamefont {I.}~\bibnamefont {Kassal}}, \bibinfo {author}
  {\bibfnamefont {J.~D.}\ \bibnamefont {Biamonte}}, \bibinfo {author}
  {\bibfnamefont {M.}~\bibnamefont {Mohseni}}, \bibinfo {author} {\bibfnamefont
  {B.~J.}\ \bibnamefont {Powell}}, \bibinfo {author} {\bibfnamefont
  {M.}~\bibnamefont {Barbieri}}, \bibinfo {author} {\bibfnamefont
  {A.}~\bibnamefont {Aspuru-Guzik}}, \ and\ \bibinfo {author} {\bibfnamefont
  {A.~G.}\ \bibnamefont {White}},\ }\href {\doibase 10.1038/nchem.483}
  {\bibfield  {journal} {\bibinfo  {journal} {Nat. Chem}\ }\textbf {\bibinfo
  {volume} {2}},\ \bibinfo {pages} {106} (\bibinfo {year} {2010})}\BibitemShut
  {NoStop}%
\bibitem [{\citenamefont {Du}\ \emph {et~al.}(2010)\citenamefont {Du},
  \citenamefont {Xu}, \citenamefont {Peng}, \citenamefont {Wang}, \citenamefont
  {Wu},\ and\ \citenamefont {Lu}}]{Du2010}%
  \BibitemOpen
  \bibfield  {author} {\bibinfo {author} {\bibfnamefont {J.}~\bibnamefont
  {Du}}, \bibinfo {author} {\bibfnamefont {N.}~\bibnamefont {Xu}}, \bibinfo
  {author} {\bibfnamefont {X.}~\bibnamefont {Peng}}, \bibinfo {author}
  {\bibfnamefont {P.}~\bibnamefont {Wang}}, \bibinfo {author} {\bibfnamefont
  {S.}~\bibnamefont {Wu}}, \ and\ \bibinfo {author} {\bibfnamefont
  {D.}~\bibnamefont {Lu}},\ }\href {\doibase 10.1103/PhysRevLett.104.030502}
  {\bibfield  {journal} {\bibinfo  {journal} {Phys. Rev. Lett.}\ }\textbf
  {\bibinfo {volume} {104}},\ \bibinfo {pages} {030502} (\bibinfo {year}
  {2010})}\BibitemShut {NoStop}%
\bibitem [{\citenamefont {Wang}\ \emph {et~al.}(2015)\citenamefont {Wang},
  \citenamefont {Dolde}, \citenamefont {Biamonte}, \citenamefont {Babbush},
  \citenamefont {Bergholm}, \citenamefont {Yang}, \citenamefont {Jakobi},
  \citenamefont {Neumann}, \citenamefont {Aspuru-Guzik}, \citenamefont
  {Whitfield},\ and\ \citenamefont {Wrachtrup}}]{Wang_quantum_2015}%
  \BibitemOpen
  \bibfield  {author} {\bibinfo {author} {\bibfnamefont {Y.}~\bibnamefont
  {Wang}}, \bibinfo {author} {\bibfnamefont {F.}~\bibnamefont {Dolde}},
  \bibinfo {author} {\bibfnamefont {J.}~\bibnamefont {Biamonte}}, \bibinfo
  {author} {\bibfnamefont {R.}~\bibnamefont {Babbush}}, \bibinfo {author}
  {\bibfnamefont {V.}~\bibnamefont {Bergholm}}, \bibinfo {author}
  {\bibfnamefont {S.}~\bibnamefont {Yang}}, \bibinfo {author} {\bibfnamefont
  {I.}~\bibnamefont {Jakobi}}, \bibinfo {author} {\bibfnamefont
  {P.}~\bibnamefont {Neumann}}, \bibinfo {author} {\bibfnamefont
  {A.}~\bibnamefont {Aspuru-Guzik}}, \bibinfo {author} {\bibfnamefont {J.~D.}\
  \bibnamefont {Whitfield}}, \ and\ \bibinfo {author} {\bibfnamefont
  {J.}~\bibnamefont {Wrachtrup}},\ }\href {\doibase 10.1021/acsnano.5b01651}
  {\bibfield  {journal} {\bibinfo  {journal} {ACS Nano}\ }\textbf {\bibinfo
  {volume} {9}},\ \bibinfo {pages} {7769} (\bibinfo {year} {2015})},\ \bibinfo
  {note} {pMID: 25905564}\BibitemShut {NoStop}%
\bibitem [{\citenamefont {Mueck}(2015)}]{Mueck2015}%
  \BibitemOpen
  \bibfield  {author} {\bibinfo {author} {\bibfnamefont {L.}~\bibnamefont
  {Mueck}},\ }\href {http://dx.doi.org/10.1038/nchem.2248} {\bibfield
  {journal} {\bibinfo  {journal} {Nat Chem}\ }\textbf {\bibinfo {volume} {7}},\
  \bibinfo {pages} {361 EP } (\bibinfo {year} {2015})}\BibitemShut {NoStop}%
\bibitem [{\citenamefont {Reiher}\ \emph {et~al.}(2017)\citenamefont {Reiher},
  \citenamefont {Wiebe}, \citenamefont {Svore}, \citenamefont {Wecker},\ and\
  \citenamefont {Troyer}}]{reiher_elucidating_2017}%
  \BibitemOpen
  \bibfield  {author} {\bibinfo {author} {\bibfnamefont {M.}~\bibnamefont
  {Reiher}}, \bibinfo {author} {\bibfnamefont {N.}~\bibnamefont {Wiebe}},
  \bibinfo {author} {\bibfnamefont {K.~M.}\ \bibnamefont {Svore}}, \bibinfo
  {author} {\bibfnamefont {D.}~\bibnamefont {Wecker}}, \ and\ \bibinfo {author}
  {\bibfnamefont {M.}~\bibnamefont {Troyer}},\ }\href {\doibase
  10.1073/pnas.1619152114} {\bibfield  {journal} {\bibinfo  {journal} {PNAS}\
  }\textbf {\bibinfo {volume} {114}},\ \bibinfo {pages} {7555} (\bibinfo {year}
  {2017})}\BibitemShut {NoStop}%
\bibitem [{\citenamefont {Shen}\ \emph {et~al.}(2017)\citenamefont {Shen},
  \citenamefont {Zhang}, \citenamefont {Zhang}, \citenamefont {Zhang},
  \citenamefont {Yung},\ and\ \citenamefont {Kim}}]{shen_quantum_2017}%
  \BibitemOpen
  \bibfield  {author} {\bibinfo {author} {\bibfnamefont {Y.}~\bibnamefont
  {Shen}}, \bibinfo {author} {\bibfnamefont {X.}~\bibnamefont {Zhang}},
  \bibinfo {author} {\bibfnamefont {S.}~\bibnamefont {Zhang}}, \bibinfo
  {author} {\bibfnamefont {J.-N.}\ \bibnamefont {Zhang}}, \bibinfo {author}
  {\bibfnamefont {M.-H.}\ \bibnamefont {Yung}}, \ and\ \bibinfo {author}
  {\bibfnamefont {K.}~\bibnamefont {Kim}},\ }\href {\doibase
  10.1103/PhysRevA.95.020501} {\bibfield  {journal} {\bibinfo  {journal} {Phys.
  Rev. A}\ }\textbf {\bibinfo {volume} {95}},\ \bibinfo {pages} {020501}
  (\bibinfo {year} {2017})}\BibitemShut {NoStop}%
\bibitem [{\citenamefont {Peruzzo}\ \emph {et~al.}(2014)\citenamefont
  {Peruzzo}, \citenamefont {McClean}, \citenamefont {Shadbolt}, \citenamefont
  {Yung}, \citenamefont {Zhou}, \citenamefont {Love}, \citenamefont
  {Aspuru-Guzik},\ and\ \citenamefont {O'Brien}}]{peruzzo_variational_2014}%
  \BibitemOpen
  \bibfield  {author} {\bibinfo {author} {\bibfnamefont {A.}~\bibnamefont
  {Peruzzo}}, \bibinfo {author} {\bibfnamefont {J.}~\bibnamefont {McClean}},
  \bibinfo {author} {\bibfnamefont {P.}~\bibnamefont {Shadbolt}}, \bibinfo
  {author} {\bibfnamefont {M.-H.}\ \bibnamefont {Yung}}, \bibinfo {author}
  {\bibfnamefont {X.-Q.}\ \bibnamefont {Zhou}}, \bibinfo {author}
  {\bibfnamefont {P.~J.}\ \bibnamefont {Love}}, \bibinfo {author}
  {\bibfnamefont {A.}~\bibnamefont {Aspuru-Guzik}}, \ and\ \bibinfo {author}
  {\bibfnamefont {J.~L.}\ \bibnamefont {O'Brien}},\ }\href {\doibase
  10.1038/ncomms5213} {\bibfield  {journal} {\bibinfo  {journal} {Nat Commun}\
  }\textbf {\bibinfo {volume} {5}} (\bibinfo {year} {2014}),\
  10.1038/ncomms5213}\BibitemShut {NoStop}%
\bibitem [{\citenamefont {Yung}\ \emph {et~al.}(2014)\citenamefont {Yung},
  \citenamefont {Casanova}, \citenamefont {Mezzacapo}, \citenamefont {McClean},
  \citenamefont {Lamata}, \citenamefont {Aspuru-Guzik},\ and\ \citenamefont
  {Solano}}]{Yung2014}%
  \BibitemOpen
  \bibfield  {author} {\bibinfo {author} {\bibfnamefont {M.~H.}\ \bibnamefont
  {Yung}}, \bibinfo {author} {\bibfnamefont {J.}~\bibnamefont {Casanova}},
  \bibinfo {author} {\bibfnamefont {A.}~\bibnamefont {Mezzacapo}}, \bibinfo
  {author} {\bibfnamefont {J.}~\bibnamefont {McClean}}, \bibinfo {author}
  {\bibfnamefont {L.}~\bibnamefont {Lamata}}, \bibinfo {author} {\bibfnamefont
  {A.}~\bibnamefont {Aspuru-Guzik}}, \ and\ \bibinfo {author} {\bibfnamefont
  {E.}~\bibnamefont {Solano}},\ }\href {http://dx.doi.org/10.1038/srep03589}
  {\bibfield  {journal} {\bibinfo  {journal} {Scientific Reports}\ }\textbf
  {\bibinfo {volume} {4}},\ \bibinfo {pages} {3589 EP } (\bibinfo {year}
  {2014})}\BibitemShut {NoStop}%
\bibitem [{\citenamefont {Barends}\ \emph {et~al.}(2015)\citenamefont
  {Barends}, \citenamefont {Lamata}, \citenamefont {Kelly}, \citenamefont
  {Garc{\'\i}a-{\'A}lvarez}, \citenamefont {Fowler}, \citenamefont {Megrant},
  \citenamefont {Jeffrey}, \citenamefont {White}, \citenamefont {Sank},
  \citenamefont {Mutus}, \citenamefont {Campbell}, \citenamefont {Chen},
  \citenamefont {Chen}, \citenamefont {Chiaro}, \citenamefont {Dunsworth},
  \citenamefont {Hoi}, \citenamefont {Neill}, \citenamefont {O'Malley},
  \citenamefont {Quintana}, \citenamefont {Roushan}, \citenamefont
  {Vainsencher}, \citenamefont {Wenner}, \citenamefont {Solano},\ and\
  \citenamefont {Martinis}}]{barends_digital_2015}%
  \BibitemOpen
  \bibfield  {author} {\bibinfo {author} {\bibfnamefont {R.}~\bibnamefont
  {Barends}}, \bibinfo {author} {\bibfnamefont {L.}~\bibnamefont {Lamata}},
  \bibinfo {author} {\bibfnamefont {J.}~\bibnamefont {Kelly}}, \bibinfo
  {author} {\bibfnamefont {L.}~\bibnamefont {Garc{\'\i}a-{\'A}lvarez}},
  \bibinfo {author} {\bibfnamefont {A.~G.}\ \bibnamefont {Fowler}}, \bibinfo
  {author} {\bibfnamefont {A.}~\bibnamefont {Megrant}}, \bibinfo {author}
  {\bibfnamefont {E.}~\bibnamefont {Jeffrey}}, \bibinfo {author} {\bibfnamefont
  {T.~C.}\ \bibnamefont {White}}, \bibinfo {author} {\bibfnamefont
  {D.}~\bibnamefont {Sank}}, \bibinfo {author} {\bibfnamefont {J.~Y.}\
  \bibnamefont {Mutus}}, \bibinfo {author} {\bibfnamefont {B.}~\bibnamefont
  {Campbell}}, \bibinfo {author} {\bibfnamefont {Y.}~\bibnamefont {Chen}},
  \bibinfo {author} {\bibfnamefont {Z.}~\bibnamefont {Chen}}, \bibinfo {author}
  {\bibfnamefont {B.}~\bibnamefont {Chiaro}}, \bibinfo {author} {\bibfnamefont
  {A.}~\bibnamefont {Dunsworth}}, \bibinfo {author} {\bibfnamefont {I.-C.}\
  \bibnamefont {Hoi}}, \bibinfo {author} {\bibfnamefont {C.}~\bibnamefont
  {Neill}}, \bibinfo {author} {\bibfnamefont {P.~J.~J.}\ \bibnamefont
  {O'Malley}}, \bibinfo {author} {\bibfnamefont {C.}~\bibnamefont {Quintana}},
  \bibinfo {author} {\bibfnamefont {P.}~\bibnamefont {Roushan}}, \bibinfo
  {author} {\bibfnamefont {A.}~\bibnamefont {Vainsencher}}, \bibinfo {author}
  {\bibfnamefont {J.}~\bibnamefont {Wenner}}, \bibinfo {author} {\bibfnamefont
  {E.}~\bibnamefont {Solano}}, \ and\ \bibinfo {author} {\bibfnamefont {J.~M.}\
  \bibnamefont {Martinis}},\ }\href {\doibase 10.1038/ncomms8654} {\bibfield
  {journal} {\bibinfo  {journal} {Nat Commun}\ }\textbf {\bibinfo {volume}
  {6}},\ \bibinfo {pages} {7654} (\bibinfo {year} {2015})}\BibitemShut
  {NoStop}%
\bibitem [{\citenamefont {McClean}\ \emph {et~al.}(2016)\citenamefont
  {McClean}, \citenamefont {Romero}, \citenamefont {Babbush},\ and\
  \citenamefont {Aspuru-Guzik}}]{mcclean_theory_2016}%
  \BibitemOpen
  \bibfield  {author} {\bibinfo {author} {\bibfnamefont {J.~R.}\ \bibnamefont
  {McClean}}, \bibinfo {author} {\bibfnamefont {J.}~\bibnamefont {Romero}},
  \bibinfo {author} {\bibfnamefont {R.}~\bibnamefont {Babbush}}, \ and\
  \bibinfo {author} {\bibfnamefont {A.}~\bibnamefont {Aspuru-Guzik}},\ }\href
  {\doibase 10.1088/1367-2630/18/2/023023} {\bibfield  {journal} {\bibinfo
  {journal} {New J. Phys.}\ }\textbf {\bibinfo {volume} {18}},\ \bibinfo
  {pages} {023023} (\bibinfo {year} {2016})}\BibitemShut {NoStop}%
\bibitem [{\citenamefont {Wang}\ \emph {et~al.}(2018)\citenamefont {Wang},
  \citenamefont {Higgott},\ and\ \citenamefont {Brierley}}]{wang18}%
  \BibitemOpen
  \bibfield  {author} {\bibinfo {author} {\bibfnamefont {D.}~\bibnamefont
  {Wang}}, \bibinfo {author} {\bibfnamefont {O.}~\bibnamefont {Higgott}}, \
  and\ \bibinfo {author} {\bibfnamefont {S.}~\bibnamefont {Brierley}},\ }\href
  {http://arxiv.org/pdf/1802.00171v1} {\bibfield  {journal} {\bibinfo
  {journal} {arxiv.org/pdf/1802.00171v1}\ } (\bibinfo {year}
  {2018})}\BibitemShut {NoStop}%
\bibitem [{\citenamefont {Feynman}(1965)}]{Feynman1965}%
  \BibitemOpen
  \bibfield  {author} {\bibinfo {author} {\bibfnamefont {R.~P.}\ \bibnamefont
  {Feynman}},\ }\href@noop {} {\emph {\bibinfo {title} {The Feynman lectures on
  physics}}}\ (\bibinfo  {publisher} {Addison-Wesley Pub. Co.},\ \bibinfo
  {year} {1963-1965})\BibitemShut {NoStop}%
\bibitem [{\citenamefont {Feynman}(1982)}]{Feynman1982}%
  \BibitemOpen
  \bibfield  {author} {\bibinfo {author} {\bibfnamefont {R.~P.}\ \bibnamefont
  {Feynman}},\ }\href@noop {} {\bibfield  {journal} {\bibinfo  {journal} {Int.
  J. Theor. Phys.}\ }\textbf {\bibinfo {volume} {21}} (\bibinfo {year}
  {1982})}\BibitemShut {NoStop}%
\bibitem [{\citenamefont {Rubin}\ \emph {et~al.}(2018)\citenamefont {Rubin},
  \citenamefont {Babbush},\ and\ \citenamefont {McClean}}]{Rubin2018}%
  \BibitemOpen
  \bibfield  {author} {\bibinfo {author} {\bibfnamefont {N.}~\bibnamefont
  {Rubin}}, \bibinfo {author} {\bibfnamefont {R.}~\bibnamefont {Babbush}}, \
  and\ \bibinfo {author} {\bibfnamefont {J.}~\bibnamefont {McClean}},\ }\href
  {https://arxiv.org/abs/1801.03524} {\bibfield  {journal} {\bibinfo  {journal}
  {arXiv:1801.03524}\ } (\bibinfo {year} {2018})}\BibitemShut {NoStop}%
\bibitem [{\citenamefont {Kivlichan}\ \emph {et~al.}(2018)\citenamefont
  {Kivlichan}, \citenamefont {McClean}, \citenamefont {Wiebe}, \citenamefont
  {Gidney}, \citenamefont {Aspuru-Guzik}, \citenamefont {Chan},\ and\
  \citenamefont {Babbush}}]{Kivlichan2018}%
  \BibitemOpen
  \bibfield  {author} {\bibinfo {author} {\bibfnamefont {I.~D.}\ \bibnamefont
  {Kivlichan}}, \bibinfo {author} {\bibfnamefont {J.}~\bibnamefont {McClean}},
  \bibinfo {author} {\bibfnamefont {N.}~\bibnamefont {Wiebe}}, \bibinfo
  {author} {\bibfnamefont {C.}~\bibnamefont {Gidney}}, \bibinfo {author}
  {\bibfnamefont {A.}~\bibnamefont {Aspuru-Guzik}}, \bibinfo {author}
  {\bibfnamefont {G.~K.-L.}\ \bibnamefont {Chan}}, \ and\ \bibinfo {author}
  {\bibfnamefont {R.}~\bibnamefont {Babbush}},\ }\href {\doibase
  10.1103/PhysRevLett.120.110501} {\bibfield  {journal} {\bibinfo  {journal}
  {Phys. Rev. Lett.}\ }\textbf {\bibinfo {volume} {120}},\ \bibinfo {pages}
  {110501} (\bibinfo {year} {2018})}\BibitemShut {NoStop}%
\bibitem [{\citenamefont {Babbush}\ \emph {et~al.}(2018)\citenamefont
  {Babbush}, \citenamefont {Wiebe}, \citenamefont {McClean}, \citenamefont
  {McClain}, \citenamefont {Neven},\ and\ \citenamefont {Chan}}]{Babbush2018}%
  \BibitemOpen
  \bibfield  {author} {\bibinfo {author} {\bibfnamefont {R.}~\bibnamefont
  {Babbush}}, \bibinfo {author} {\bibfnamefont {N.}~\bibnamefont {Wiebe}},
  \bibinfo {author} {\bibfnamefont {J.}~\bibnamefont {McClean}}, \bibinfo
  {author} {\bibfnamefont {J.}~\bibnamefont {McClain}}, \bibinfo {author}
  {\bibfnamefont {H.}~\bibnamefont {Neven}}, \ and\ \bibinfo {author}
  {\bibfnamefont {G.~K.-L.}\ \bibnamefont {Chan}},\ }\href {\doibase
  10.1103/PhysRevX.8.011044} {\bibfield  {journal} {\bibinfo  {journal} {Phys.
  Rev. X}\ }\textbf {\bibinfo {volume} {8}},\ \bibinfo {pages} {011044}
  (\bibinfo {year} {2018})}\BibitemShut {NoStop}%
\bibitem [{\citenamefont {Sherrill}\ and\ \citenamefont
  {Schaefer}(1999)}]{Sherrill1999}%
  \BibitemOpen
  \bibfield  {author} {\bibinfo {author} {\bibfnamefont {C.~D.}\ \bibnamefont
  {Sherrill}}\ and\ \bibinfo {author} {\bibfnamefont {H.~F.}\ \bibnamefont
  {Schaefer}}\ }(\bibinfo  {publisher} {Academic Press},\ \bibinfo {year}
  {1999})\ pp.\ \bibinfo {pages} {143 -- 269}\BibitemShut {NoStop}%
\bibitem [{\citenamefont {Bartlett}\ and\ \citenamefont
  {Musia\l{}}(2007)}]{Bartlett2007}%
  \BibitemOpen
  \bibfield  {author} {\bibinfo {author} {\bibfnamefont {R.~J.}\ \bibnamefont
  {Bartlett}}\ and\ \bibinfo {author} {\bibfnamefont {M.}~\bibnamefont
  {Musia\l{}}},\ }\href {\doibase 10.1103/RevModPhys.79.291} {\bibfield
  {journal} {\bibinfo  {journal} {Rev. Mod. Phys.}\ }\textbf {\bibinfo {volume}
  {79}},\ \bibinfo {pages} {291} (\bibinfo {year} {2007})}\BibitemShut
  {NoStop}%
\bibitem [{Note1()}]{Note1}%
  \BibitemOpen
  \bibinfo {note} {The correct scaling will be $\begin {pmatrix} N_{b} \\
  N_{\protect \rm el}\end {pmatrix}$, where $N_{b}$ is the number of basis
  functions and $N_{\protect \rm el}$ is the number of electrons.}\BibitemShut
  {Stop}%
\bibitem [{\citenamefont {Raghavachari}\ \emph {et~al.}(1989)\citenamefont
  {Raghavachari}, \citenamefont {Trucks}, \citenamefont {Pople},\ and\
  \citenamefont {Head-Gordon}}]{Raghavachari1989}%
  \BibitemOpen
  \bibfield  {author} {\bibinfo {author} {\bibfnamefont {K.}~\bibnamefont
  {Raghavachari}}, \bibinfo {author} {\bibfnamefont {G.~W.}\ \bibnamefont
  {Trucks}}, \bibinfo {author} {\bibfnamefont {J.~A.}\ \bibnamefont {Pople}}, \
  and\ \bibinfo {author} {\bibfnamefont {M.}~\bibnamefont {Head-Gordon}},\
  }\href {\doibase https://doi.org/10.1016/S0009-2614(89)87395-6} {\bibfield
  {journal} {\bibinfo  {journal} {Chem Phys Lett}\ }\textbf {\bibinfo {volume}
  {157}},\ \bibinfo {pages} {479 } (\bibinfo {year} {1989})}\BibitemShut
  {NoStop}%
\bibitem [{\citenamefont {Becke}(2013)}]{Becke2013}%
  \BibitemOpen
  \bibfield  {author} {\bibinfo {author} {\bibfnamefont {A.~D.}\ \bibnamefont
  {Becke}},\ }\href {\doibase 10.1063/1.4790598} {\bibfield  {journal}
  {\bibinfo  {journal} {J Chem Phys}\ }\textbf {\bibinfo {volume} {138}},\
  \bibinfo {pages} {074109} (\bibinfo {year} {2013})}\BibitemShut {NoStop}%
\bibitem [{\citenamefont {Ziesche}\ \emph {et~al.}(1997)\citenamefont
  {Ziesche}, \citenamefont {Gunnarsson}, \citenamefont {John},\ and\
  \citenamefont {Beck}}]{Ziesche1997}%
  \BibitemOpen
  \bibfield  {author} {\bibinfo {author} {\bibfnamefont {P.}~\bibnamefont
  {Ziesche}}, \bibinfo {author} {\bibfnamefont {O.}~\bibnamefont {Gunnarsson}},
  \bibinfo {author} {\bibfnamefont {W.}~\bibnamefont {John}}, \ and\ \bibinfo
  {author} {\bibfnamefont {H.}~\bibnamefont {Beck}},\ }\href {\doibase
  10.1103/PhysRevB.55.10270} {\bibfield  {journal} {\bibinfo  {journal} {Phys.
  Rev. B}\ }\textbf {\bibinfo {volume} {55}},\ \bibinfo {pages} {10270}
  (\bibinfo {year} {1997})}\BibitemShut {NoStop}%
\bibitem [{\citenamefont {Jordan}\ and\ \citenamefont
  {Wigner}(1928)}]{jordan_uber_1928}%
  \BibitemOpen
  \bibfield  {author} {\bibinfo {author} {\bibfnamefont {P.}~\bibnamefont
  {Jordan}}\ and\ \bibinfo {author} {\bibfnamefont {E.}~\bibnamefont
  {Wigner}},\ }\href {\doibase 10.1007/BF01331938} {\bibfield  {journal}
  {\bibinfo  {journal} {Zeitschrift f\"ur Physik}\ }\textbf {\bibinfo {volume}
  {47}},\ \bibinfo {pages} {631} (\bibinfo {year} {1928})}\BibitemShut
  {NoStop}%
\bibitem [{\citenamefont {Bravyi}\ and\ \citenamefont
  {Kitaev}(2002)}]{bravyi_fermionic_2002}%
  \BibitemOpen
  \bibfield  {author} {\bibinfo {author} {\bibfnamefont {S.~B.}\ \bibnamefont
  {Bravyi}}\ and\ \bibinfo {author} {\bibfnamefont {A.~Y.}\ \bibnamefont
  {Kitaev}},\ }\href {\doibase 10.1006/aphy.2002.6254} {\bibfield  {journal}
  {\bibinfo  {journal} {Annals of Physics}\ }\textbf {\bibinfo {volume}
  {298}},\ \bibinfo {pages} {210} (\bibinfo {year} {2002})}\BibitemShut
  {NoStop}%
\bibitem [{\citenamefont {Bravyi}\ \emph {et~al.}(2017)\citenamefont {Bravyi},
  \citenamefont {Gambetta}, \citenamefont {Mezzacapo},\ and\ \citenamefont
  {Temme}}]{bravyi_tapering_2017}%
  \BibitemOpen
  \bibfield  {author} {\bibinfo {author} {\bibfnamefont {S.}~\bibnamefont
  {Bravyi}}, \bibinfo {author} {\bibfnamefont {J.~M.}\ \bibnamefont
  {Gambetta}}, \bibinfo {author} {\bibfnamefont {A.}~\bibnamefont {Mezzacapo}},
  \ and\ \bibinfo {author} {\bibfnamefont {K.}~\bibnamefont {Temme}},\ }\href
  {http://arxiv.org/abs/1701.08213} {\bibfield  {journal} {\bibinfo  {journal}
  {arXiv:1701.08213 [quant-ph]}\ } (\bibinfo {year} {2017})}\BibitemShut
  {NoStop}%
\bibitem [{\citenamefont {O'Malley}\ \emph {et~al.}(2016)\citenamefont
  {O'Malley}, \citenamefont {Babbush}, \citenamefont {Kivlichan}, \citenamefont
  {Romero}, \citenamefont {McClean}, \citenamefont {Barends}, \citenamefont
  {Kelly}, \citenamefont {Roushan}, \citenamefont {Tranter}, \citenamefont
  {Ding}, \citenamefont {Campbell}, \citenamefont {Chen}, \citenamefont {Chen},
  \citenamefont {Chiaro}, \citenamefont {Dunsworth}, \citenamefont {Fowler},
  \citenamefont {Jeffrey}, \citenamefont {Lucero}, \citenamefont {Megrant},
  \citenamefont {Mutus}, \citenamefont {Neeley}, \citenamefont {Neill},
  \citenamefont {Quintana}, \citenamefont {Sank}, \citenamefont {Vainsencher},
  \citenamefont {Wenner}, \citenamefont {White}, \citenamefont {Coveney},
  \citenamefont {Love}, \citenamefont {Neven}, \citenamefont {Aspuru-Guzik},\
  and\ \citenamefont {Martinis}}]{omalley_scalable_2016}%
  \BibitemOpen
  \bibfield  {author} {\bibinfo {author} {\bibfnamefont {P.}~\bibnamefont
  {O'Malley}}, \bibinfo {author} {\bibfnamefont {R.}~\bibnamefont {Babbush}},
  \bibinfo {author} {\bibfnamefont {I.}~\bibnamefont {Kivlichan}}, \bibinfo
  {author} {\bibfnamefont {J.}~\bibnamefont {Romero}}, \bibinfo {author}
  {\bibfnamefont {J.}~\bibnamefont {McClean}}, \bibinfo {author} {\bibfnamefont
  {R.}~\bibnamefont {Barends}}, \bibinfo {author} {\bibfnamefont
  {J.}~\bibnamefont {Kelly}}, \bibinfo {author} {\bibfnamefont
  {P.}~\bibnamefont {Roushan}}, \bibinfo {author} {\bibfnamefont
  {A.}~\bibnamefont {Tranter}}, \bibinfo {author} {\bibfnamefont
  {N.}~\bibnamefont {Ding}}, \bibinfo {author} {\bibfnamefont {B.}~\bibnamefont
  {Campbell}}, \bibinfo {author} {\bibfnamefont {Y.}~\bibnamefont {Chen}},
  \bibinfo {author} {\bibfnamefont {Z.}~\bibnamefont {Chen}}, \bibinfo {author}
  {\bibfnamefont {B.}~\bibnamefont {Chiaro}}, \bibinfo {author} {\bibfnamefont
  {A.}~\bibnamefont {Dunsworth}}, \bibinfo {author} {\bibfnamefont
  {A.}~\bibnamefont {Fowler}}, \bibinfo {author} {\bibfnamefont
  {E.}~\bibnamefont {Jeffrey}}, \bibinfo {author} {\bibfnamefont
  {E.}~\bibnamefont {Lucero}}, \bibinfo {author} {\bibfnamefont
  {A.}~\bibnamefont {Megrant}}, \bibinfo {author} {\bibfnamefont
  {J.}~\bibnamefont {Mutus}}, \bibinfo {author} {\bibfnamefont
  {M.}~\bibnamefont {Neeley}}, \bibinfo {author} {\bibfnamefont
  {C.}~\bibnamefont {Neill}}, \bibinfo {author} {\bibfnamefont
  {C.}~\bibnamefont {Quintana}}, \bibinfo {author} {\bibfnamefont
  {D.}~\bibnamefont {Sank}}, \bibinfo {author} {\bibfnamefont {A.}~\bibnamefont
  {Vainsencher}}, \bibinfo {author} {\bibfnamefont {J.}~\bibnamefont {Wenner}},
  \bibinfo {author} {\bibfnamefont {T.}~\bibnamefont {White}}, \bibinfo
  {author} {\bibfnamefont {P.}~\bibnamefont {Coveney}}, \bibinfo {author}
  {\bibfnamefont {P.}~\bibnamefont {Love}}, \bibinfo {author} {\bibfnamefont
  {H.}~\bibnamefont {Neven}}, \bibinfo {author} {\bibfnamefont
  {A.}~\bibnamefont {Aspuru-Guzik}}, \ and\ \bibinfo {author} {\bibfnamefont
  {J.}~\bibnamefont {Martinis}},\ }\href {\doibase 10.1103/PhysRevX.6.031007}
  {\bibfield  {journal} {\bibinfo  {journal} {Phys. Rev. X}\ }\textbf {\bibinfo
  {volume} {6}},\ \bibinfo {pages} {031007} (\bibinfo {year}
  {2016})}\BibitemShut {NoStop}%
\bibitem [{\citenamefont {Kandala}\ \emph {et~al.}(2017)\citenamefont
  {Kandala}, \citenamefont {Mezzacapo}, \citenamefont {Temme}, \citenamefont
  {Takita}, \citenamefont {Brink}, \citenamefont {Chow},\ and\ \citenamefont
  {Gambetta}}]{kandala_hardware-efficient_2017}%
  \BibitemOpen
  \bibfield  {author} {\bibinfo {author} {\bibfnamefont {A.}~\bibnamefont
  {Kandala}}, \bibinfo {author} {\bibfnamefont {A.}~\bibnamefont {Mezzacapo}},
  \bibinfo {author} {\bibfnamefont {K.}~\bibnamefont {Temme}}, \bibinfo
  {author} {\bibfnamefont {M.}~\bibnamefont {Takita}}, \bibinfo {author}
  {\bibfnamefont {M.}~\bibnamefont {Brink}}, \bibinfo {author} {\bibfnamefont
  {J.~M.}\ \bibnamefont {Chow}}, \ and\ \bibinfo {author} {\bibfnamefont
  {J.~M.}\ \bibnamefont {Gambetta}},\ }\href {\doibase 10.1038/nature23879}
  {\bibfield  {journal} {\bibinfo  {journal} {Nature}\ }\textbf {\bibinfo
  {volume} {549}},\ \bibinfo {pages} {242} (\bibinfo {year}
  {2017})}\BibitemShut {NoStop}%
\bibitem [{\citenamefont {Colless}\ \emph {et~al.}(2018)\citenamefont
  {Colless}, \citenamefont {Ramasesh}, \citenamefont {Dahlen}, \citenamefont
  {Blok}, \citenamefont {Kimchi-Schwartz}, \citenamefont {McClean},
  \citenamefont {Carter}, \citenamefont {de~Jong},\ and\ \citenamefont
  {Siddiqi}}]{Colless2018}%
  \BibitemOpen
  \bibfield  {author} {\bibinfo {author} {\bibfnamefont {J.~I.}\ \bibnamefont
  {Colless}}, \bibinfo {author} {\bibfnamefont {V.~V.}\ \bibnamefont
  {Ramasesh}}, \bibinfo {author} {\bibfnamefont {D.}~\bibnamefont {Dahlen}},
  \bibinfo {author} {\bibfnamefont {M.~S.}\ \bibnamefont {Blok}}, \bibinfo
  {author} {\bibfnamefont {M.~E.}\ \bibnamefont {Kimchi-Schwartz}}, \bibinfo
  {author} {\bibfnamefont {J.~R.}\ \bibnamefont {McClean}}, \bibinfo {author}
  {\bibfnamefont {J.}~\bibnamefont {Carter}}, \bibinfo {author} {\bibfnamefont
  {W.~A.}\ \bibnamefont {de~Jong}}, \ and\ \bibinfo {author} {\bibfnamefont
  {I.}~\bibnamefont {Siddiqi}},\ }\href {\doibase 10.1103/PhysRevX.8.011021}
  {\bibfield  {journal} {\bibinfo  {journal} {Phys. Rev. X}\ }\textbf {\bibinfo
  {volume} {8}},\ \bibinfo {pages} {011021} (\bibinfo {year}
  {2018})}\BibitemShut {NoStop}%
\bibitem [{\citenamefont {Hempel}\ \emph {et~al.}(2018)\citenamefont {Hempel},
  \citenamefont {Maier}, \citenamefont {Romero}, \citenamefont {McClean},
  \citenamefont {Monz}, \citenamefont {Shen}, \citenamefont {Jurcevic},
  \citenamefont {Lanyon}, \citenamefont {Love}, \citenamefont {Babbush},
  \citenamefont {Aspuru-Guzik}, \citenamefont {Blatt},\ and\ \citenamefont
  {Roos}}]{Hempel2018}%
  \BibitemOpen
  \bibfield  {author} {\bibinfo {author} {\bibfnamefont {C.}~\bibnamefont
  {Hempel}}, \bibinfo {author} {\bibfnamefont {C.}~\bibnamefont {Maier}},
  \bibinfo {author} {\bibfnamefont {J.}~\bibnamefont {Romero}}, \bibinfo
  {author} {\bibfnamefont {J.}~\bibnamefont {McClean}}, \bibinfo {author}
  {\bibfnamefont {T.}~\bibnamefont {Monz}}, \bibinfo {author} {\bibfnamefont
  {H.}~\bibnamefont {Shen}}, \bibinfo {author} {\bibfnamefont {P.}~\bibnamefont
  {Jurcevic}}, \bibinfo {author} {\bibfnamefont {B.}~\bibnamefont {Lanyon}},
  \bibinfo {author} {\bibfnamefont {P.}~\bibnamefont {Love}}, \bibinfo {author}
  {\bibfnamefont {R.}~\bibnamefont {Babbush}}, \bibinfo {author} {\bibfnamefont
  {A.}~\bibnamefont {Aspuru-Guzik}}, \bibinfo {author} {\bibfnamefont
  {R.}~\bibnamefont {Blatt}}, \ and\ \bibinfo {author} {\bibfnamefont
  {C.}~\bibnamefont {Roos}},\ }\href {https://arxiv.org/abs/1803.10238}
  {\bibfield  {journal} {\bibinfo  {journal} {arXiv:1803.10238}\ } (\bibinfo
  {year} {2018})}\BibitemShut {NoStop}%
\bibitem [{\citenamefont {Moll}\ \emph {et~al.}(2016)\citenamefont {Moll},
  \citenamefont {Fuhrer}, \citenamefont {Staar},\ and\ \citenamefont
  {Tavernelli}}]{moll_optimizing_2016}%
  \BibitemOpen
  \bibfield  {author} {\bibinfo {author} {\bibfnamefont {N.}~\bibnamefont
  {Moll}}, \bibinfo {author} {\bibfnamefont {A.}~\bibnamefont {Fuhrer}},
  \bibinfo {author} {\bibfnamefont {P.}~\bibnamefont {Staar}}, \ and\ \bibinfo
  {author} {\bibfnamefont {I.}~\bibnamefont {Tavernelli}},\ }\href {\doibase
  10.1088/1751-8113/49/29/295301} {\bibfield  {journal} {\bibinfo  {journal}
  {J. Phys. A: Math. Theor.}\ }\textbf {\bibinfo {volume} {49}},\ \bibinfo
  {pages} {295301} (\bibinfo {year} {2016})}\BibitemShut {NoStop}%
\bibitem [{\citenamefont {Whitfield}\ \emph {et~al.}(2012)\citenamefont
  {Whitfield}, \citenamefont {Love},\ and\ \citenamefont
  {Aspuru-Guzik}}]{whitfield_computational_2012}%
  \BibitemOpen
  \bibfield  {author} {\bibinfo {author} {\bibfnamefont {J.~D.}\ \bibnamefont
  {Whitfield}}, \bibinfo {author} {\bibfnamefont {P.~J.}\ \bibnamefont {Love}},
  \ and\ \bibinfo {author} {\bibfnamefont {A.}~\bibnamefont {Aspuru-Guzik}},\
  }\href {\doibase 10.1039/C2CP42695A} {\bibfield  {journal} {\bibinfo
  {journal} {Phys. Chem. Chem. Phys.}\ }\textbf {\bibinfo {volume} {15}},\
  \bibinfo {pages} {397} (\bibinfo {year} {2012})}\BibitemShut {NoStop}%
\bibitem [{\citenamefont {Seeley}\ \emph {et~al.}(2012)\citenamefont {Seeley},
  \citenamefont {Richard},\ and\ \citenamefont {Love}}]{Seeley2012}%
  \BibitemOpen
  \bibfield  {author} {\bibinfo {author} {\bibfnamefont {J.~T.}\ \bibnamefont
  {Seeley}}, \bibinfo {author} {\bibfnamefont {M.~J.}\ \bibnamefont {Richard}},
  \ and\ \bibinfo {author} {\bibfnamefont {P.~J.}\ \bibnamefont {Love}},\
  }\href@noop {} {\bibfield  {journal} {\bibinfo  {journal} {J. Chem. Phys}\
  }\textbf {\bibinfo {volume} {137}},\ \bibinfo {pages} {224109} (\bibinfo
  {year} {2012})}\BibitemShut {NoStop}%
\bibitem [{\citenamefont {Romero}\ \emph {et~al.}(2017)\citenamefont {Romero},
  \citenamefont {Babbush}, \citenamefont {McClean}, \citenamefont {Hempel},
  \citenamefont {Love},\ and\ \citenamefont
  {Aspuru-Guzik}}]{romero_strategies_2017}%
  \BibitemOpen
  \bibfield  {author} {\bibinfo {author} {\bibfnamefont {J.}~\bibnamefont
  {Romero}}, \bibinfo {author} {\bibfnamefont {R.}~\bibnamefont {Babbush}},
  \bibinfo {author} {\bibfnamefont {J.~R.}\ \bibnamefont {McClean}}, \bibinfo
  {author} {\bibfnamefont {C.}~\bibnamefont {Hempel}}, \bibinfo {author}
  {\bibfnamefont {P.}~\bibnamefont {Love}}, \ and\ \bibinfo {author}
  {\bibfnamefont {A.}~\bibnamefont {Aspuru-Guzik}},\ }\href
  {http://arxiv.org/abs/1701.02691} {\bibfield  {journal} {\bibinfo  {journal}
  {arXiv:1701.02691 [quant-ph]}\ } (\bibinfo {year} {2017})}\BibitemShut
  {NoStop}%
\bibitem [{\citenamefont {Spall}(2000)}]{spall_adaptive_2000}%
  \BibitemOpen
  \bibfield  {author} {\bibinfo {author} {\bibfnamefont {J.}~\bibnamefont
  {Spall}},\ }\href {\doibase 10.1109/TAC.2000.880982} {\bibfield  {journal}
  {\bibinfo  {journal} {IEEE Transactions on Automatic Control}\ }\textbf
  {\bibinfo {volume} {45}},\ \bibinfo {pages} {1839} (\bibinfo {year}
  {2000})}\BibitemShut {NoStop}%
\bibitem [{\citenamefont {Fetter}\ and\ \citenamefont
  {Walecka}(2003)}]{Fetter2003}%
  \BibitemOpen
  \bibfield  {author} {\bibinfo {author} {\bibfnamefont {A.~L.}\ \bibnamefont
  {Fetter}}\ and\ \bibinfo {author} {\bibfnamefont {J.~D.}\ \bibnamefont
  {Walecka}},\ }\href@noop {} {\emph {\bibinfo {title} {Quantum Theory of
  Many-particle Systems}}}\ (\bibinfo  {publisher} {Courier Corporation},\
  \bibinfo {year} {2003})\BibitemShut {NoStop}%
\bibitem [{\citenamefont {Szabo}\ and\ \citenamefont
  {Ostlund}(2012)}]{szabo12}%
  \BibitemOpen
  \bibfield  {author} {\bibinfo {author} {\bibfnamefont {A.}~\bibnamefont
  {Szabo}}\ and\ \bibinfo {author} {\bibfnamefont {N.~S.}\ \bibnamefont
  {Ostlund}},\ }\href {http://subhh.eblib.com/patron/FullRecord.aspx?p=1894806}
  {\emph {\bibinfo {title} {Modern Quantum Chemistry: Introduction to Advanced
  Electronic Structure Theory}}},\ Dover Books on Chemistry\ (\bibinfo
  {publisher} {{Dover Publications}},\ \bibinfo {address} {Newburyport},\
  \bibinfo {year} {2012})\BibitemShut {NoStop}%
\bibitem [{Note2()}]{Note2}%
  \BibitemOpen
  \bibinfo {note} {To implement the UCCSD wavefunction Ansatz, we expand the
  Hamiltonian in the basis function of the occupied and virtual HF orbitals,
  with a number of occupied orbitals equal to the number of electrons in the
  system. This picture has the advantage of allowing a simple interpretation of
  the expansion of the reference wavefunction in terms of excited
  configurations (Slater determinants) Further manipulations of the molecular
  Hamiltonians in the unmodified second quantized form (Eq.~\protect \textup
  {\hbox {\mathsurround \z@ \protect \normalfont (\ignorespaces \ref
  {Eq:H_sec_quant}\unskip \@@italiccorr )}}) or in the p/h formulation can be
  used to further reduce the number of required qubits. One possibility, is to
  apply the projection scheme introduced in~\cite {moll_optimizing_2016,
  Barkoutsos_fermionic_2017}, which allow to restrict the search space from the
  entire Fock space to the sector of the Hilbert space with the selected number
  of electrons. However, this procedure will make the physical interpretation
  of the UCC expansion less evident and the mapping to the quantum circuits
  more cumbersome. For these reasons, in this work we will restrict to the
  simplest map that encodes each basis function in a different
  qubit.}\BibitemShut {Stop}%
\bibitem [{\citenamefont {Bartlett}\ \emph {et~al.}(1989)\citenamefont
  {Bartlett}, \citenamefont {Kucharski},\ and\ \citenamefont
  {Noga}}]{Bartlett1989}%
  \BibitemOpen
  \bibfield  {author} {\bibinfo {author} {\bibfnamefont {R.~J.}\ \bibnamefont
  {Bartlett}}, \bibinfo {author} {\bibfnamefont {S.~A.}\ \bibnamefont
  {Kucharski}}, \ and\ \bibinfo {author} {\bibfnamefont {J.}~\bibnamefont
  {Noga}},\ }\href {\doibase https://doi.org/10.1016/S0009-2614(89)87372-5}
  {\bibfield  {journal} {\bibinfo  {journal} {Chem Phys Lett}\ }\textbf
  {\bibinfo {volume} {155}},\ \bibinfo {pages} {133 } (\bibinfo {year}
  {1989})}\BibitemShut {NoStop}%
\bibitem [{\citenamefont {Kutzelnigg}(1991)}]{Kutzelnigg1991}%
  \BibitemOpen
  \bibfield  {author} {\bibinfo {author} {\bibfnamefont {W.}~\bibnamefont
  {Kutzelnigg}},\ }\href {\doibase 10.1007/BF01117418} {\bibfield  {journal}
  {\bibinfo  {journal} {Theor chimica acta}\ }\textbf {\bibinfo {volume}
  {80}},\ \bibinfo {pages} {349} (\bibinfo {year} {1991})}\BibitemShut
  {NoStop}%
\bibitem [{\citenamefont {Harsha}\ \emph {et~al.}(2018)\citenamefont {Harsha},
  \citenamefont {Shiozaki},\ and\ \citenamefont {Scuseria}}]{Harsha2018}%
  \BibitemOpen
  \bibfield  {author} {\bibinfo {author} {\bibfnamefont {G.}~\bibnamefont
  {Harsha}}, \bibinfo {author} {\bibfnamefont {T.}~\bibnamefont {Shiozaki}}, \
  and\ \bibinfo {author} {\bibfnamefont {G.~E.}\ \bibnamefont {Scuseria}},\
  }\href {\doibase 10.1063/1.5011033} {\bibfield  {journal} {\bibinfo
  {journal} {J Chem Phys}\ }\textbf {\bibinfo {volume} {148}},\ \bibinfo
  {pages} {044107} (\bibinfo {year} {2018})}\BibitemShut {NoStop}%
\bibitem [{qis()}]{qiskit_webpage}%
  \BibitemOpen
  \href {https://www.qiskit.org/} {\emph {\bibinfo {title} {QISKit Open Source
  Quantum Information Software Kit - https://www.qiskit.org/}}}\BibitemShut
  {NoStop}%
\bibitem [{\citenamefont {McKay}\ \emph {et~al.}(2016)\citenamefont {McKay},
  \citenamefont {Filipp}, \citenamefont {Mezzacapo}, \citenamefont {Magesan},
  \citenamefont {Chow},\ and\ \citenamefont {Gambetta}}]{McKay2016}%
  \BibitemOpen
  \bibfield  {author} {\bibinfo {author} {\bibfnamefont {D.~C.}\ \bibnamefont
  {McKay}}, \bibinfo {author} {\bibfnamefont {S.}~\bibnamefont {Filipp}},
  \bibinfo {author} {\bibfnamefont {A.}~\bibnamefont {Mezzacapo}}, \bibinfo
  {author} {\bibfnamefont {E.}~\bibnamefont {Magesan}}, \bibinfo {author}
  {\bibfnamefont {J.~M.}\ \bibnamefont {Chow}}, \ and\ \bibinfo {author}
  {\bibfnamefont {J.~M.}\ \bibnamefont {Gambetta}},\ }\href {\doibase
  10.1103/PhysRevApplied.6.064007} {\bibfield  {journal} {\bibinfo  {journal}
  {Phys. Rev. Applied}\ }\textbf {\bibinfo {volume} {6}},\ \bibinfo {pages}
  {064007} (\bibinfo {year} {2016})}\BibitemShut {NoStop}%
\bibitem [{\citenamefont {Roth}\ \emph {et~al.}(2017)\citenamefont {Roth},
  \citenamefont {Ganzhorn}, \citenamefont {Moll}, \citenamefont {Filipp},
  \citenamefont {Salis},\ and\ \citenamefont {Schmidt}}]{Roth2017}%
  \BibitemOpen
  \bibfield  {author} {\bibinfo {author} {\bibfnamefont {M.}~\bibnamefont
  {Roth}}, \bibinfo {author} {\bibfnamefont {M.}~\bibnamefont {Ganzhorn}},
  \bibinfo {author} {\bibfnamefont {N.}~\bibnamefont {Moll}}, \bibinfo {author}
  {\bibfnamefont {S.}~\bibnamefont {Filipp}}, \bibinfo {author} {\bibfnamefont
  {G.}~\bibnamefont {Salis}}, \ and\ \bibinfo {author} {\bibfnamefont
  {S.}~\bibnamefont {Schmidt}},\ }\href {\doibase 10.1103/PhysRevA.96.062323}
  {\bibfield  {journal} {\bibinfo  {journal} {Phys. Rev. A}\ }\textbf {\bibinfo
  {volume} {96}},\ \bibinfo {pages} {062323} (\bibinfo {year}
  {2017})}\BibitemShut {NoStop}%
\bibitem [{\citenamefont {Egger}\ \emph {et~al.}(2018)\citenamefont {Egger},
  \citenamefont {Ganzhorn}, \citenamefont {Fuhrer}, \citenamefont {Mueller},
  \citenamefont {Barkoutsos}, \citenamefont {Moll}, \citenamefont
  {Tavernelli},\ and\ \citenamefont {Filipp}}]{Egger2018}%
  \BibitemOpen
  \bibfield  {author} {\bibinfo {author} {\bibfnamefont {D.~J.}\ \bibnamefont
  {Egger}}, \bibinfo {author} {\bibfnamefont {G.}~\bibnamefont {Ganzhorn},
  \bibfnamefont {Marc amd~Salis}}, \bibinfo {author} {\bibfnamefont
  {A.}~\bibnamefont {Fuhrer}}, \bibinfo {author} {\bibfnamefont
  {P.}~\bibnamefont {Mueller}}, \bibinfo {author} {\bibfnamefont {P.~K.}\
  \bibnamefont {Barkoutsos}}, \bibinfo {author} {\bibfnamefont
  {N.}~\bibnamefont {Moll}}, \bibinfo {author} {\bibfnamefont {I.}~\bibnamefont
  {Tavernelli}}, \ and\ \bibinfo {author} {\bibfnamefont {S.}~\bibnamefont
  {Filipp}},\ }\href@noop {} {\bibfield  {journal} {\bibinfo  {journal}
  {arXiv:1804.04900 [quant-ph]}\ } (\bibinfo {year} {2018})}\BibitemShut
  {NoStop}%
\bibitem [{\citenamefont {McClean}\ \emph {et~al.}(2018)\citenamefont
  {McClean}, \citenamefont {Boixo}, \citenamefont {Smelyanskiy}, \citenamefont
  {Babbush},\ and\ \citenamefont {Neven}}]{McClean_2018_Baren}%
  \BibitemOpen
  \bibfield  {author} {\bibinfo {author} {\bibfnamefont {J.}~\bibnamefont
  {McClean}}, \bibinfo {author} {\bibfnamefont {S.}~\bibnamefont {Boixo}},
  \bibinfo {author} {\bibfnamefont {V.}~\bibnamefont {Smelyanskiy}}, \bibinfo
  {author} {\bibfnamefont {R.}~\bibnamefont {Babbush}}, \ and\ \bibinfo
  {author} {\bibfnamefont {H.}~\bibnamefont {Neven}},\ }\href@noop {}
  {\bibfield  {journal} {\bibinfo  {journal} {arXiv:1803.11173v1 [quant-ph]}\ }
  (\bibinfo {year} {2018})}\BibitemShut {NoStop}%
\bibitem [{\citenamefont {Murnaghan}(1962)}]{murnaghan62}%
  \BibitemOpen
  \bibfield  {author} {\bibinfo {author} {\bibfnamefont {F.~D.}\ \bibnamefont
  {Murnaghan}},\ }\href@noop {} {\emph {\bibinfo {title} {The unitary and
  rotation groups}}},\ \bibinfo {series} {Lectures on applied mathematics},
  Vol.~\bibinfo {volume} {3}\ (\bibinfo  {publisher} {{Spartan Books}},\
  \bibinfo {address} {Washington, D.C.},\ \bibinfo {year} {1962})\BibitemShut
  {NoStop}%
\bibitem [{\citenamefont {Alkauskas}\ \emph {et~al.}(2004)\citenamefont
  {Alkauskas}, \citenamefont {Baratoff},\ and\ \citenamefont
  {Bruder}}]{Alkauskas2004}%
  \BibitemOpen
  \bibfield  {author} {\bibinfo {author} {\bibfnamefont {A.}~\bibnamefont
  {Alkauskas}}, \bibinfo {author} {\bibfnamefont {A.}~\bibnamefont {Baratoff}},
  \ and\ \bibinfo {author} {\bibfnamefont {C.}~\bibnamefont {Bruder}},\
  }\href@noop {} {\bibfield  {journal} {\bibinfo  {journal} {J Phys Chem A}\
  }\textbf {\bibinfo {volume} {108}},\ \bibinfo {pages} {6863} (\bibinfo {year}
  {2004})}\BibitemShut {NoStop}%
\bibitem [{\citenamefont {Jensen}(2007)}]{Jensen2007}%
  \BibitemOpen
  \bibfield  {author} {\bibinfo {author} {\bibfnamefont {F.}~\bibnamefont
  {Jensen}},\ }\href@noop {} {\emph {\bibinfo {title} {Introduction to
  Computational Chemistry}}}\ (\bibinfo  {publisher} {John Wiley and Sons
  Ltd},\ \bibinfo {year} {2007})\BibitemShut {NoStop}%
\bibitem [{\citenamefont {Unruh}(1995)}]{unruh_maintaining_1995}%
  \BibitemOpen
  \bibfield  {author} {\bibinfo {author} {\bibfnamefont {W.~G.}\ \bibnamefont
  {Unruh}},\ }\href {\doibase 10.1103/PhysRevA.51.992} {\bibfield  {journal}
  {\bibinfo  {journal} {Phys. Rev. A}\ }\textbf {\bibinfo {volume} {51}},\
  \bibinfo {pages} {992} (\bibinfo {year} {1995})}\BibitemShut {NoStop}%
\bibitem [{\citenamefont {Nielsen}(2005)}]{nielsen05}%
  \BibitemOpen
  \bibfield  {author} {\bibinfo {author} {\bibfnamefont {M.~A.}\ \bibnamefont
  {Nielsen}},\ }\href@noop {} {\emph {\bibinfo {title} {The Fermionic canonical
  commutation relations and the Jordan-Wigner transform}}},\ \bibinfo {type}
  {Tech. Rep.}\ (\bibinfo  {institution} {University of Queensland},\ \bibinfo
  {year} {2005})\BibitemShut {NoStop}%
\bibitem [{\citenamefont {Fletcher}(1970)}]{fletcher70}%
  \BibitemOpen
  \bibfield  {author} {\bibinfo {author} {\bibfnamefont {R.}~\bibnamefont
  {Fletcher}},\ }\href {\doibase 10.1093/comjnl/13.3.317} {\bibfield  {journal}
  {\bibinfo  {journal} {Comput J}\ }\textbf {\bibinfo {volume} {13}},\ \bibinfo
  {pages} {317} (\bibinfo {year} {1970})}\BibitemShut {NoStop}%
\bibitem [{\citenamefont {Ditchfield}\ \emph {et~al.}(1971)\citenamefont
  {Ditchfield}, \citenamefont {Hehre},\ and\ \citenamefont
  {Pople}}]{Ditchfield1971}%
  \BibitemOpen
  \bibfield  {author} {\bibinfo {author} {\bibfnamefont {R.}~\bibnamefont
  {Ditchfield}}, \bibinfo {author} {\bibfnamefont {W.~J.}\ \bibnamefont
  {Hehre}}, \ and\ \bibinfo {author} {\bibfnamefont {J.~A.}\ \bibnamefont
  {Pople}},\ }\href {\doibase 10.1063/1.1674902} {\bibfield  {journal}
  {\bibinfo  {journal} {The Journal of Chemical Physics}\ }\textbf {\bibinfo
  {volume} {54}},\ \bibinfo {pages} {724} (\bibinfo {year} {1971})},\ \Eprint
  {http://arxiv.org/abs/https://doi.org/10.1063/1.1674902}
  {https://doi.org/10.1063/1.1674902} \BibitemShut {NoStop}%
\bibitem [{\citenamefont {Gunnarsson}\ and\ \citenamefont
  {Lundqvist}(1976)}]{Gunnarsson1976}%
  \BibitemOpen
  \bibfield  {author} {\bibinfo {author} {\bibfnamefont {O.}~\bibnamefont
  {Gunnarsson}}\ and\ \bibinfo {author} {\bibfnamefont {B.~I.}\ \bibnamefont
  {Lundqvist}},\ }\href {\doibase 10.1103/PhysRevB.13.4274} {\bibfield
  {journal} {\bibinfo  {journal} {Phys. Rev. B}\ }\textbf {\bibinfo {volume}
  {13}},\ \bibinfo {pages} {4274} (\bibinfo {year} {1976})}\BibitemShut
  {NoStop}%
\bibitem [{\citenamefont {Wecker}\ \emph {et~al.}(2015)\citenamefont {Wecker},
  \citenamefont {Hastings},\ and\ \citenamefont {Troyer}}]{Wecker2015a}%
  \BibitemOpen
  \bibfield  {author} {\bibinfo {author} {\bibfnamefont {D.}~\bibnamefont
  {Wecker}}, \bibinfo {author} {\bibfnamefont {M.~B.}\ \bibnamefont
  {Hastings}}, \ and\ \bibinfo {author} {\bibfnamefont {M.}~\bibnamefont
  {Troyer}},\ }\href {\doibase 10.1103/PhysRevA.92.042303} {\bibfield
  {journal} {\bibinfo  {journal} {Phys. Rev. A}\ }\textbf {\bibinfo {volume}
  {92}},\ \bibinfo {pages} {042303} (\bibinfo {year} {2015})}\BibitemShut
  {NoStop}%
\bibitem [{\citenamefont {Barkoutsos}\ \emph {et~al.}(2017)\citenamefont
  {Barkoutsos}, \citenamefont {Moll}, \citenamefont {Staar}, \citenamefont
  {Mueller}, \citenamefont {Fuhrer}, \citenamefont {Filipp}, \citenamefont
  {Troyer},\ and\ \citenamefont {Tavernelli}}]{Barkoutsos_fermionic_2017}%
  \BibitemOpen
  \bibfield  {author} {\bibinfo {author} {\bibfnamefont {P.~K.}\ \bibnamefont
  {Barkoutsos}}, \bibinfo {author} {\bibfnamefont {N.}~\bibnamefont {Moll}},
  \bibinfo {author} {\bibfnamefont {P.}~\bibnamefont {Staar}}, \bibinfo
  {author} {\bibfnamefont {P.}~\bibnamefont {Mueller}}, \bibinfo {author}
  {\bibfnamefont {A.}~\bibnamefont {Fuhrer}}, \bibinfo {author} {\bibfnamefont
  {S.}~\bibnamefont {Filipp}}, \bibinfo {author} {\bibfnamefont
  {M.}~\bibnamefont {Troyer}}, \ and\ \bibinfo {author} {\bibfnamefont
  {I.}~\bibnamefont {Tavernelli}},\ }\href@noop {} {\bibfield  {journal}
  {\bibinfo  {journal} {arXiv:1706.03637}\ } (\bibinfo {year}
  {2017})}\BibitemShut {NoStop}%
\bibitem [{\citenamefont {Barenco}\ \emph {et~al.}(1995)\citenamefont
  {Barenco}, \citenamefont {Bennett}, \citenamefont {Cleve}, \citenamefont
  {DiVincenzo}, \citenamefont {Margolus}, \citenamefont {Shor}, \citenamefont
  {Sleator}, \citenamefont {Smolin},\ and\ \citenamefont
  {Weinfurter}}]{Barenco1995}%
  \BibitemOpen
  \bibfield  {author} {\bibinfo {author} {\bibfnamefont {A.}~\bibnamefont
  {Barenco}}, \bibinfo {author} {\bibfnamefont {C.~H.}\ \bibnamefont
  {Bennett}}, \bibinfo {author} {\bibfnamefont {R.}~\bibnamefont {Cleve}},
  \bibinfo {author} {\bibfnamefont {D.~P.}\ \bibnamefont {DiVincenzo}},
  \bibinfo {author} {\bibfnamefont {N.}~\bibnamefont {Margolus}}, \bibinfo
  {author} {\bibfnamefont {P.}~\bibnamefont {Shor}}, \bibinfo {author}
  {\bibfnamefont {T.}~\bibnamefont {Sleator}}, \bibinfo {author} {\bibfnamefont
  {J.~A.}\ \bibnamefont {Smolin}}, \ and\ \bibinfo {author} {\bibfnamefont
  {H.}~\bibnamefont {Weinfurter}},\ }\href {\doibase 10.1103/PhysRevA.52.3457}
  {\bibfield  {journal} {\bibinfo  {journal} {Phys. Rev. A}\ }\textbf {\bibinfo
  {volume} {52}},\ \bibinfo {pages} {3457} (\bibinfo {year}
  {1995})}\BibitemShut {NoStop}%
\end{thebibliography}%

\appendix

\renewcommand{\theequation}{A\thesection.\arabic{equation}}
\setcounter{equation}{0}

\section{Approximations in UCCSD}
\label{section:appendix_UCCSD_approx}
The q-UCC expansion operator for the HF ground state $|\Phi_0 \rangle$ is given by 
\begin{equation}
\hat{U}(\vec{\theta)} = e^{\hat{T}(\vec{\theta})-\hat{T}^{\dagger}(\vec{\theta})}
\label{eq:U}
\end{equation}
With the excitation operator $\hat T(\vec{\theta})=\hat T_{(1)}(\vec{\theta})+\hat T_{(2)}(\vec{\theta})$ restricted to the single $\hat T_{(1)}(\vec{\theta})$ and double $\hat T_{(2)}(\vec{\theta})$ excitations, Eq. \eqref{eq:U} becomes
\begin{equation}
\hat{U}(\vec{\theta}) = e^{\hat{T}_{(1)}(\vec{\theta})+\hat{T}_{(2)}(\vec{\theta}) -\hat{T}^{\dagger}_{(1)}(\vec{\theta})-\hat{T}^{\dagger}_{(2)}(\vec{\theta})} \, .
\label{eq:op}
\end{equation}
Substituting $\hat{T}_{(1)}(\vec{\theta}) = \sum_{ij} \theta_{ij} \hat{a}_i^{\dagger} \hat{a}_j$
and $\hat{T}_{(2)}(\vec{\theta}) = \sum_{ijkl} \theta_{ijkl} \hat{a}^{\dagger}_i \hat{a}^{\dagger}_j \hat{a}_k \hat{a}_l$, 
Eq. \eqref{eq:op} reads

\begin{align}
\hat{U} ( \vec{\theta} ) = {\rm{exp}} \Bigg( & \sum_{ij}\theta_{ij} \hat{a}^{\dagger}_i \hat{a}_j   + \sum_{ijkl} \theta_{ijkl} \hat{a}_i^{\dagger} \hat{a}_j^{\dagger}  \hat{a}_k \hat{a}_l \nonumber \\
& - \sum_{ij} \theta_{ij }\hat{a}_j^{\dagger} \hat{a}_i  - \sum_{ijkl} \hat{a}_l^{\dagger} \hat{a}_k^{\dagger} \hat{a}_j \hat{a}_i \Bigg)
\label{eq:U_expnd}
\end{align}

with $\vec{\theta} = ( \{\theta_{ij}\},\{\theta_{ijkl}\} )$ and $\theta_{ij},\theta_{ijkl} \in \mathbb{R}$. 
Using the Trotter approximation  to the first order
$e^{\hat A+ \hat B} \approx 
e^{\hat A} e^{\hat B}$ 
with $\hat A = \hat T_{(1)}(\vec{\theta}) - \hat T_{(1)}^{\dagger}(\vec{\theta}) $ and  $\hat B = \hat T_{(2)}(\vec{\theta}) - \hat T_{(2)}^{\dagger}(\vec{\theta})$, Eq. \eqref{eq:U_expnd} becomes
\begin{align}
\hat{U}(\vec{\theta}) &= 
\exp{\left(\sum_{ij}   \theta_{ij} (\hat{a}_i^{\dagger} \hat{a}_j - \hat{a}_j^{\dagger} \hat{a}_i) \right)}  \nonumber \\
        & \times \exp{\left(\sum_{ijkl} \theta_{ijkl} (\hat{a}^{\dagger}_i \hat{a}^{\dagger}_j \hat{a}_k \hat{a}_l - \hat{a}^{\dagger}_l \hat{a}^{\dagger}_k \hat{a}_j \hat{a}_i)\right)} \, .
\label{eq:U_sums}
\end{align}
Applying once more the Trotter expansion to first order we get
\begin{equation}
\begin{split}
\hat{U}(\vec{\theta}) &= \prod_{ij} \exp{\left(\theta_{ij} (\hat{a}_i^{\dagger} \hat{a}_j - \hat{a}_j^{\dagger} \hat{a}_i) \right)} \\
                           &\times \prod_{ijkl} \exp{\left( \theta_{ijkl} (\hat{a}^{\dagger}_i \hat{a}^{\dagger}_j \hat{a}_k \hat{a}_l - \hat{a}^{\dagger}_l \hat{a}^{\dagger}_k \hat{a}_j \hat{a}_i) \right)} \, .
\end{split}
\label{eq:U_prdcts}
\end{equation}

At this point, we apply the Jordan-Wigner transformation defined by 
$\hat{a}_j=1^{\otimes j }\otimes \frac{1}{2}(\hat{X}+i\hat{Y}) \otimes \hat{Z}^{\otimes N-j-1}$ 
and
$\hat{a}^{\dagger}_j=1^{\otimes j }\otimes \frac{1}{2}(\hat{X}-i\hat{Y}) \otimes \hat{Z}^{\otimes N-j-1}$ 
with the Pauli matrices $\{ \hat{X} = \hat{\sigma}^x, \hat{Y}= \hat{\sigma}^y, \hat{Z}=\hat{\sigma}^z \}$ and $j=[0,.., N_q-1]$ where $N_q$ is the number of qubits. 
For $i>j>k>l$, without loss of generality, Eq.\eqref{eq:U_prdcts} can be expressed as

\begin{equation}
\begin{split}
\hat{U}(\vec{\theta}) &= \prod_{i>j} 
        \exp{\left(\frac{i\theta_{ij}}{2} \bigotimes_{a=j+1}^{i-1}\hat{\sigma}_{z,a}(\hat{\sigma}_{y,j} \hat{\sigma}_{x,i} - \hat{\sigma}_{x,j} \hat{\sigma}_{y,i})\right)} \\ 
        & \times  \prod_{i>j>k>l} \exp
        \Big(
          \frac{i\theta_{ijkl}}{8} \bigotimes_{b=l+1}^{k-1}\hat{\sigma}_{z,b} \bigotimes_{a=j+1}^{i-1}\hat{\sigma}_{z,a} \\
          & \quad \quad \quad \quad \quad \quad \times(\hat{\sigma}_{x,l}\hat{\sigma}_{x,k}\hat{\sigma}_{y,j}\hat{\sigma}_{x,i}   
         +\hat{\sigma}_{y,l}\hat{\sigma}_{x,k}\hat{\sigma}_{y,j}\hat{\sigma}_{y,i} \\
         & \quad \quad \quad \quad \quad \quad
         + \hat{\sigma}_{x,l}\hat{\sigma}_{y,k}\hat{\sigma}_{y,j}\hat{\sigma}_{y,i}  +\hat{\sigma}_{x,l}\hat{\sigma}_{x,k}\hat{\sigma}_{x,j}\hat{\sigma}_{y,i} \\ & \quad \quad \quad \quad \quad \quad             -\hat{\sigma}_{y,l}\hat{\sigma}_{x,k}\hat{\sigma}_{x,j}\hat{\sigma}_{x,i}  -\hat{\sigma}_{x,l}\hat{\sigma}_{y,k}\hat{\sigma}_{x,j}\hat{\sigma}_{x,i}\\
        & \quad \quad \quad \quad \quad \quad
        -\hat{\sigma}_{y,l}\hat{\sigma}_{y,k}\hat{\sigma}_{y,j}\hat{\sigma}_{x,i}  -\hat{\sigma}_{y,l}\hat{\sigma}_{y,k}\hat{\sigma}_{x,j}\hat{\sigma}_{y,i}) 
                \Big) \, .
\end{split}
\label{eq:U_pauli}
\end{equation}
Using the assignment $\hat{A} = \hat{\sigma}_{y,j} \hat{\sigma}_{x,i}$ and $\hat{B}=-\hat{\sigma}_{x,j} \hat{\sigma}_{y,i}$ 
all commutators $[\hat{A},[\hat{A},\hat{B}]]$, $[\hat{B},[\hat{A},\hat{B}]]$ and $[\hat{A},\hat{B}]$ vanish. Therefore, the Glauber's formula $e^{ \hat{A}+\hat B}=e^{\hat A}e^{\hat B}e^{\frac{1}{2}[\hat A,\hat B]}$ simplifies exactly 
to $e^{\hat A+\hat B}=e^{\hat A}e^{\hat B}$ (similarly fo the double excitation terms). Therefore, the q-UCCSD expansion operator can finally be written as 

\begin{equation}
\begin{split}
\hat{U}(\vec{\theta}) & = \prod_{i>j} \exp{\left(\frac{i\theta_{ij}}{2} \bigotimes_{a=j+1}^{i-1}\hat{\sigma}_{z,a} (\hat{\sigma}_{y,j} \hat{\sigma}_{x,i})\right)} \, \\ &\times \exp{\left(- \frac{i\theta_{ij}}{2} \bigotimes_{a=j+1}^{i-1}\hat{\sigma}_{z,a} (\hat{\sigma}_{x,j} \hat{\sigma}_{y,i})\right)} \\
                          & \times \prod_{i>j>k>l} \exp{ \Big(\frac{i\theta_{ijkl}}{8} \bigotimes_{b=l+1}^{k-1}\hat{\sigma}_{z,b} \bigotimes_{a=j+1}^{i-1}\hat{\sigma}_{z,a} (\hat{\sigma}_{x,l}\hat{\sigma}_{x,k}\hat{\sigma}_{y,j}\hat{\sigma}_{x,i})\Big)} \\   
                          & \times \exp{ \Big(\frac{i\theta_{ijkl}}{8} \bigotimes_{b=l+1}^{k-1}\hat{\sigma}_{z,b} \bigotimes_{a=j+1}^{i-1}\hat{\sigma}_{z,a} (\hat{\sigma}_{y,l}\hat{\sigma}_{x,k}\hat{\sigma}_{y,j}\hat{\sigma}_{y,i})\Big)} \\ & \times \exp{ \Big(\frac{i\theta_{ijkl}}{8} \bigotimes_{b=l+1}^{k-1}\hat{\sigma}_{z,b} \bigotimes_{a=j+1}^{i-1}\hat{\sigma}_{z,a} (\hat{\sigma}_{x,l}\hat{\sigma}_{y,k}\hat{\sigma}_{y,j}\hat{\sigma}_{y,i})\Big)} \\ 
                          & \times \exp{ \Big(\frac{i\theta_{ijkl}}{8} \bigotimes_{b=l+1}^{k-1}\hat{\sigma}_{z,b} \bigotimes_{a=j+1}^{i-1}\hat{\sigma}_{z,a} (\hat{\sigma}_{x,l}\hat{\sigma}_{x,k}\hat{\sigma}_{x,j}\hat{\sigma}_{y,i})\Big)} \\ 
                          & 
                          \times \exp{\Big(-\frac{i\theta_{ijkl}}{8} \bigotimes_{b=l+1}^{k-1}\hat{\sigma}_{z,b} \bigotimes_{a=j+1}^{i-1}\hat{\sigma}_{z,a}(\hat{\sigma}_{y,l}\hat{\sigma}_{x,k}\hat{\sigma}_{x,j}\hat{\sigma}_{x,i})\Big)} \\ 
                          & 
                            \times \exp{ \Big(-\frac{i\theta_{ijkl}}{8} \bigotimes_{b=l+1}^{k-1}\hat{\sigma}_{z,b} \bigotimes_{a=j+1}^{i-1}\hat{\sigma}_{z,a} (\hat{\sigma}_{x,l}\hat{\sigma}_{y,k}\hat{\sigma}_{x,j}\hat{\sigma}_{x,i})\Big)} \\ & \times \exp{ \Big(-\frac{i\theta_{ijkl}}{8} \bigotimes_{b=l+1}^{k-1}\hat{\sigma}_{z,b} \bigotimes_{a=j+1}^{i-1}\hat{\sigma}_{z,a} ( \hat{\sigma}_{y,l}\hat{\sigma}_{y,k}\hat{\sigma}_{y,j}\hat{\sigma}_{x,i})\Big)} \\ & \times \exp{ \Big(-\frac{i\theta_{ijkl}}{8} \bigotimes_{b=l+1}^{k-1}\hat{\sigma}_{z,b} \bigotimes_{a=j+1}^{i-1}\hat{\sigma}_{z,a} (\hat{\sigma}_{y,l}\hat{\sigma}_{y,k}\hat{\sigma}_{x,j}\hat{\sigma}_{y,i})\Big)}
                          \, .
\end{split}
\label{eq:U_final}
\end{equation}

In a more compact way $\hat{U}(\vec{\theta}) =  \prod_{i>j} \hat{U}_{ij} \prod_{i>j>k>l} \hat{U}_{ijkl}$, notice that its parts $[\hat{U}_{ij},\hat{U}_{i'j'}] \neq 0$, $[\hat{U}_{ij},\hat{U}_{i'j'k'l'}] \neq 0$ and 
$[\hat{U}_{ijkl},\hat{U}_{i'j'k'l'}] \neq 0$ when sets $\{i,j,k,l\}$, $\{i',j',k',l'\}$ share same indices (e.g. $i\neq i',j\neq j', k \neq k', l=l'$).

\section{Definition of the gate operations spanning multiple qubits}
\label{App:boxes}

In the case of the operation spanning multiple qubits 
(see dashed boxes in Figs.~\ref{fig:UCCSD_circ_T1}
and ~\ref{fig:circuits}),
we used the following schemes (assuming nearest-neighbor connectivity)

\begin{figure}[h!]
    \centering
    \includegraphics[width = 0.95\columnwidth]{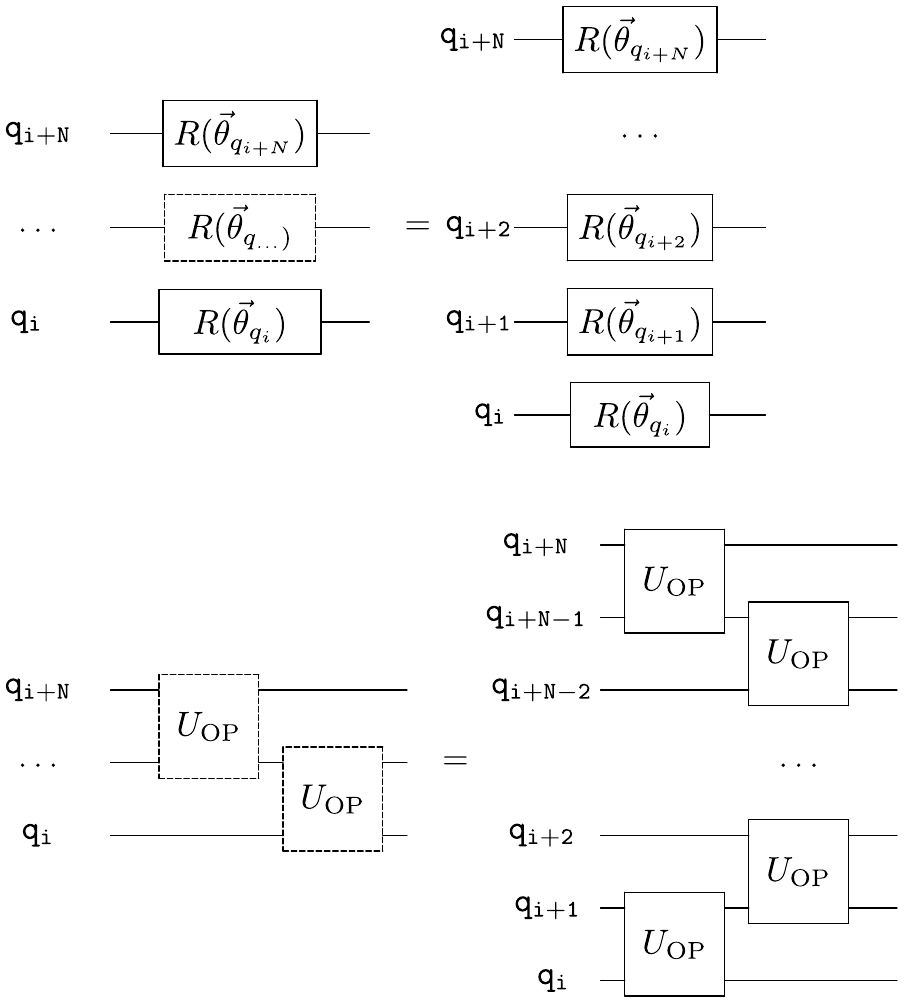}
    \label{fig:Appendix_1}
    \caption{Definition of single- and two-qubit gate blocks (dashed lines) that span multiple qubits.}
\end{figure}
The top circuit describes the decomposition of the composite one-qubit operation (dashed box in the l.h.s.) into a sequence of one-qubit operations between the starting qubit, $q_i$ and the final one, $q_{i+N}$, each one parametrized by a different set of angles. A similar procedure applies to the two-qubit operations as shown in the lower panel. 
The operator $U_{\text{OP}}$ stands for one of the operators discussed in Section~\ref{sss:Heuristic}.

\section{Decomposition of the exchange gates in elementary gates}
\label{App:Decomposition}

\begin{figure}[h!]
    \centering
    \includegraphics[width = 0.95\columnwidth]{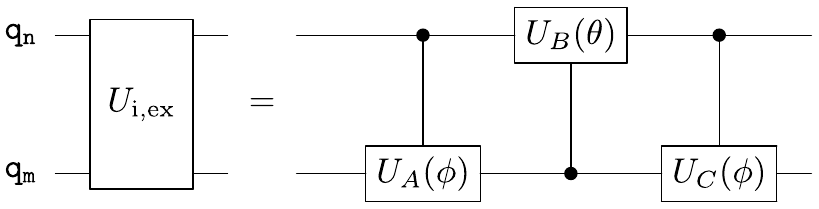}
    \caption{Decomposition of an exchange gate ($i=1,2$) between qubit $n$ and $m$ into their elementary gates.}
    \label{fig:Decomposition}
\end{figure}

The two exchange gates $U_{\rm{1,ex}}$ and $U_{\rm{2,ex}}$ can be directly implemented in a single step using the approach outlined in~\cite{McKay2016,Roth2017, Egger2018}. 
For sake of completeness, in order to emphasize the gain in gate count here we report their decomposition into elementary gates~\cite{Barenco1995}.
The result is summarized in Fig.~\eqref{fig:Decomposition} for $U_{\rm{i,ex}}$ (with $i=1,2$)

Where, for the $U_{\rm{1,ex}}$:

\[
U_{\rm{1,ex}}=
  \begin{pmatrix}
    1 & 0 & 0 & 0 \\
    0 & cos(\theta) & e^{i\phi}sin(\theta)    & 0 \\
    0 & e^{-i\phi}sin(\theta)   & -cos(\theta)  & 0 \\
    0 & 0 & 0 & 1 
  \end{pmatrix}
\]

and the $U_A$,$U_B$ and $U_C$ gates are 

\[
U_{\rm{A}}=
  \begin{pmatrix}
    0          & e^{-i\phi}  \\
    e^{-i\phi} & 0 \\
  \end{pmatrix}
\]

\[
U_{\rm{B}}=
  \begin{pmatrix}
    cos(\theta)          & sin(\theta)  \\
    sin(\theta) & -cos(\theta) \\
  \end{pmatrix}
\]

\[
U_{\rm{c}}=
  \begin{pmatrix}
    0          & e^{i\phi}  \\
    e^{i\phi} & 0
  \end{pmatrix} \, ,
\]

and for $U_{\rm{2,ex}}$:

\[
U_{\rm{2,ex}}=
  \begin{pmatrix} 
      1 & 0 & 0 & 0 \\
      0 &\cos 2 \theta & -i \sin 2 \theta & 0\\
      0 &  -i \sin 2 \theta & \cos 2 \theta & 0\\
      0 & 0 & 0 & 1
    \end{pmatrix}
\]

the $U_A$,$U_B$ and $U_C$ gates are:

\[
U_{\rm{A}}=
  \begin{pmatrix}
    0 & 1 \\
    1 & 0 \\
  \end{pmatrix}
\]

\[
U_{\rm{B}}=
  \begin{pmatrix}
    cos(2\theta)          & - i sin(2\theta)  \\
    - i sin(2\theta) & cos(2\theta) \\
  \end{pmatrix}
\]

\[
U_{\rm{c}}=
  \begin{pmatrix}
    0 & 1  \\
    1 & 0 \\
  \end{pmatrix} \, .
\]

\end{document}